\documentclass[11pt]{article}
\usepackage{amsfonts}
\usepackage{amssymb}
\usepackage{amsmath}
\usepackage{soul}
\usepackage{epsfig}
\usepackage{graphicx}
\usepackage{bm}
\usepackage{color}
\usepackage{hyperref}

\newlength{\bredde}
\def\slash#1{\settowidth{\bredde}{$#1$}\ifmmode\,\raisebox{.15ex}{/}
\hspace*{-\bredde} #1\else$\,\raisebox{.15ex}{/}\hspace*{-\bredde} #1$\fi}
\newcommand{\be}{\begin{equation}}
\newcommand{\ee}{\end{equation}}
\newcommand{\bea}{\begin{eqnarray}}
\newcommand{\eea}{\end{eqnarray}}
\newcommand{\nn}{\nonumber}
\newcommand{\eins}{\leavevmode\hbox{\small1\kern-3.8pt\normalsize1}}
\newcommand{\e}{\mbox{e}}

\newcommand{\erfc}{\mbox{erfc}}

\newcommand{\sect}[1]{\setcounter{equation}{0}\section{#1}}

\newcommand{\U}{{\rm U\,}}

\newcommand{\tr}{{\rm tr\,}}

\date{}

\begin{document}
\topmargin -1.4cm
\oddsidemargin -0.8cm
\evensidemargin -0.8cm
\title{\Large{{\bf
Universal microscopic correlation functions for products of
truncated unitary matrices
}}}

\vspace{1.5cm}

\author{~\\{\sc Gernot Akemann}$^{1}$, {\sc Zdzislaw Burda}$^{2}$,
{\sc Mario Kieburg}$^{1}$, and {\sc Taro Nagao}$^3$
\\~\\
$^1$Department of Physics,
Bielefeld University,\\
Postfach 100131,
D-33501 Bielefeld, Germany\\~\\
$^2$Marian Smoluchowski Institute of Physics, Jagellonian University,\\
Reymonta 4, 30-059 Krak\'ow, Poland\\~\\
$^3$Graduate School of Mathematics, Nagoya~University,\\
Chikusa-ku, Nagoya~464-8602, Japan\\~\\
{\it E-mail: akemann@physik.uni-bielefeld.de, zdzislaw.burda@uj.edu.pl,}\\ {\it mkieburg@physik.uni-bielefeld.de, nagao@math.nagoya-u.ac.jp}
}

\maketitle
\vfill
\begin{abstract}
We investigate the spectral properties of the product of $M$ complex non-Hermitian random matrices that are obtained by removing $L$ rows and columns of larger unitary random matrices uniformly distributed on the group $\U(N+L)$.
Such matrices are called truncated unitary matrices  or random contractions.
We first derive the joint probability distribution for the complex eigenvalues of the product matrix for fixed $N,\ L$, and $M$, given by a standard determinantal point process in the complex plane.
The weight however is non-standard and can be expressed in terms of the Meijer G-function.
The explicit knowledge of all eigenvalue correlation functions and the corresponding kernel allows us to take  various large $N$ (and $L$) limits at fixed $M$.
At strong non-unitarity, with $L/N$ finite, the eigenvalues condense on a domain inside the unit circle. At the edge and in the bulk we find the same universal microscopic kernel as for a single complex non-Hermitian matrix from the Ginibre ensemble. At the origin we find the same new universality classes labelled by $M$ as for the product of $M$ matrices from the Ginibre ensemble.
Keeping a fixed size of truncation, $L$, when $N$ goes to infinity leads to weak non-unitarity, with most eigenvalues on the unit circle as for unitary matrices. Here we find a new microscopic edge kernel that generalizes the known results for $M=1$. We briefly comment on the case when each product matrix results from a truncation of different size $L_j$.

\end{abstract}
\vfill

\thispagestyle{empty}
\newpage

\renewcommand{\thefootnote}{\arabic{footnote}}
\setcounter{footnote}{0}

\sect{Introduction}\label{intro}

The topic of products of random matrices first introduced in \cite{FKe} has seen a certain renaissance in recent years. Such products have been applied in a variety of disciplines  ranging from the problem of entanglement in quantum mechanics \cite{Collins,Zycz}, quantum chromodynamics with chemical potential \cite{Osborn}, combinatorics \cite{Karol} to finance \cite{Bouchaud}, wireless telecommunications \cite{Ralf} and image processing \cite{Andy}. See Ref.~\cite{handbook} and references therein for a general and recent overview of random matrix theory (RMT).

In this work we are interested in products of random matrices which are related to the unitary group, namely truncations thereof also called sub-unitary or random contractions.  In general, unitary matrices play an important role in time evolution in quantum mechanics or matrix valued diffusion \cite{Nowak}. Another classical problem where (sub-) unitary random matrices have been extensively used is that of chaotic scattering on mesoscopic devices, as reviewed e.g. in \cite{Carlo}. Typically the scattering process is represented by a unitary $S$-matrix, and the reflection or transmission between the leads of such a device are sub-matrices of the $S$-matrix. We refer to \cite{YanDima} for a recent review on scattering in chaotic systems using RMT. It is therefore very interesting to study also the spectral properties of such sub-unitary matrices obtained from truncating the full $S$-matrix, and such an analysis has been already made for a single random matrix in \cite{KS}. Subunitary matrices have also been considered in the context of many-body quantum states and a one-component plasma of charges on the pseudo-sphere \cite{ForK}. Previously an alternative representation of subunitary matrices was constructed in \cite{FK}. Here the concept of weak non-unitarity was introduced by adding an imaginary  finite rank perturbation to the Gaussian Unitary Ensemble, leading to the same correlations as in \cite{KS} for truncated unitary matrices, in the limit where the size of truncation is kept fixed while the matrix size goes to infinity. Truncations of orthogonal matrices have also been considered, see \cite{KS,KSZ}, which are used when the system fulfills additional symmetries (e.g. time reversal invariance).

Very recently the joint probability distribution and spectral properties of the complex eigenvalues \cite{ABu,AStr,Jesper,ARRS,peter2,IK} and the real singular values \cite{AKW,AIK,Lun,KZ} of products of $M$ random matrices of size $N\times N$ have been computed for finite $N$ and $M$. The matrices that were multiplied were drawn from the Ginibre ensemble \cite{Ginibre} of non-Hermitian matrices with a Gaussian probability distribution. Such exact results open up the possibility of a rigorous asymptotic large $N$ analysis. First steps in that direction have already been taken in these works. Prior to these finite-$N$ results it had already been observed, that when multiplying random matrices from different ensembles a large degree of universality can be seen \cite{Burda-rect}, at least regarding the limiting mean density that was derived in \cite{Ralf,IK,BJW,goetze}.

It is the purpose of this paper to extend this approach to the product of several truncated unitary random matrices. Several questions may serve as a further motivation. Can the aforementioned analytic solution for finite $N$ be extended to such ensembles of matrices, and what are the corresponding universality classes in the large $N$ limit? Second, in \cite{KS} the truncation of a unitary matrix was viewed as a mean to introduce decoherence in a quantum system. How does such a system evolve in several steps?

In order to partly answer some of these questions we consider the product of $M$ truncated (or sub-) unitary matrices $X_j$, with $j=1,\ldots,M$, which originate each from truncating a unitary matrix of size $N+L_j$ distributed with respect to the Haar measure of the unitary group $\U(N+L_j)$ by removing $L_j$ rows and columns. Obviously the resulting product matrix is non-unitary and its complex eigenvalues lie inside the unit disk.
We want to study the spectral properties of the product matrix, first at fixed $N,\ L_j$, and $M$, and then in various large $N$ limits, always keeping $M$ fixed in this work.
When writing up this paper a related paper \cite{ARRS} appeared, where among other results the same product of truncated unitary matrices is considered, for  fixed $N,\ L_j$, and $M$, only. Whenever we can compare the results from the first part of our paper they agree with their findings. Moreover very recently another work~\cite{IK} co-authored by one of the authors was published as a preprint discussing the general case of products of rectangular matrices. Those products also comprise truncated unitary matrices which were given as an example. However the derivation for those matrices was not as  thoroughly discussed as it is done here. In \cite{IK} only the macroscopic limit with $L_j\propto N\to\infty$ and fixed $M$ was presented. Neither a discussion of local fluctuations nor an analysis of universality were given in \cite{IK} or in \cite{ARRS} as it will be considered here. The focus of \cite{IK} lied on weak commutation relations of random matrices and general algebraical structures of them which were derived for all three Dyson indices ($\beta=1,2,4$) in a unifying way.

The remainder of this paper is organized as follows. In section \ref{sec2} we calculate the joint probability density of the product matrix as well as of its complex eigenvalues in subsection \ref{sec2.1}. Because it leads to a standard determinantal point process, with the eigenvalues repelling each other through the modulus square of the Vandermonde determinant, all $k$-point density correlation functions immediately follow, once the weight function is determined in subsection \ref{sec2.2}. In appendix \ref{sec2.3} we also consider a more general setting, when the individual factors $X_j$ result from truncations of random unitary matrices of different sizes, $N+L_j$, down to the dimension $N$. Section \ref{sec3} is devoted to various large $N$ limits starting with the strong non-unitary limit in subsection \ref{sec3.2}. Here $L/N$ remains fixed and positive, and various universal results are recovered for the eigenvalue correlation functions, depending on whether the fluctuations at the edge, in the bulk or at the origin are considered.
The weak non-unitarity large $N$ limit when keeping $L$ fixed is performed in subsection \ref{sec3.1}. Here we find a new class of correlation functions for $M>1$ at the edge of the unit circle.
We conclude in section \ref{sec4}. Some further technical details about the measure and the joint probability distribution are collected in two further appendices~\ref{app1} and \ref{app2}, respectively.

\sect{The solution for finite $N$}\label{sec2}

We consider the product of $M$ random matrices $X^{(M)}=X_MX_{M-1}\cdots X_1$. Each of these random matrices $X_j$ has the size $N\times N$ and is truncated from a larger $(N+L)\times (N+L)$ unitary matrix $U_j$. The unitary matrices $U_j$ are identically, independently distributed via the normalized Haar-measure on $\U(N+L)$. We recall the measure of the product matrix $X^{(M)}$ in subsection~\ref{sec2.1} and derive the joint probability density of its complex eigenvalues as well as all $k$-point density correlation functions.
In subsection~\ref{sec2.2} the weight function is determined in terms of the Meijer G-function, giving several examples. In appendix~\ref{sec2.3} we generalize this result to different truncations of the individual $X_j$,
resulting from unitary matrices of different sizes $N+L_j$ truncated to $N$, i.e. $L_j\neq L_i$.

\subsection{Probability measure and joint probability density}\label{sec2.1}

Consider a square $N\times N$ sub-block $X$ of a single unitary
random matrix $U$ distributed according to the Haar-measure on the unitary
group $\U(N+L)$ \cite{KS}
\begin{eqnarray}\label{unitaryj}
U=\left(\begin{array}{cc} X & W \\ V & Y \end{array}\right)\in\U(N+L)\ .
\end{eqnarray}
This block is a random matrix which is usually referred to as
truncated unitary random matrix, or random contraction, cf. \cite{KS,ForK,BHJ}. The probability measure
for truncated matrices with $L\geq0$ is
\begin{equation}
\begin{split}
d\mu(X) 
&\propto d[X] \int {\det}^{-L}[\imath H-\eins_N]\exp[{\tr (XX^\dagger -\eins_N)(\imath H-\eins_N)}]\  d[H],
\end{split}
\label{measurejI}
\end{equation}
where we integrate over a $N\times N$ Hermitian matrix $H$. Let $L\geq 1$ in the following, since $L=0$ corresponds to the full Haar measure of $\U(N)$ defined by a Dirac delta function, see eq.~\ref{defHaar}. In the case $L\geq N$ this integral can be readily performed,
\begin{equation}
\begin{split}
d\mu(X) 
& \propto  {\det}^{L-N}(\eins_{N}- XX^\dagger)\Theta(\eins_{N}-X^\dagger X)d[X].
\end{split}
\label{measurej}
\end{equation}
This measure is  also known as Jacobi measure \cite{petersbook}. The symbol $\Theta$ denotes
the matrix Heaviside function which is equal to one for positive definite
matrices and zero otherwise; $d[X]$ is a shorthand notation for
the flat measure $d[X] = \prod_{ij} d {\Re e}X_{ij} d {\Im m} X_{ij}$.
In appendix~\ref{app1} we briefly recall the derivation of eq.~\eqref{measurejI} and of the Jacobi measure~(\ref{measurej})
for truncated matrices.

We are interested in the product
$X\equiv X^{(M)}=X_M\cdots X_1$ of $M$
independent truncated matrices $X_j$, for $j=1,\ldots, M$.
The probability measure for such a product is
\begin{eqnarray}\label{matrixweight1}
d\nu(X)= d[X] \int \delta(X - X_M \cdots X_1) \prod\limits_{j=1}^M d\mu(X_j)\ ,
\end{eqnarray}
with $d\mu(X_j)$ given by eq.~(\ref{measurejI}). The Dirac delta function of  a complex matrix $A$ is defined as the product of the Dirac delta functions of its real independent variables, i.e. $\delta(A)=\prod_{i,j}\delta^{(2)}(A_{ij})$,
where the two-dimensional Dirac delta function for a complex variable $z$ is given as $\delta^{(2)}(z)\equiv\delta(\Re e(z))$ $\delta(\Im m (z))$.

We are going to derive
the joint probability density function for the eigenvalues of the product matrix $X$.
To this end we parameterize the measure (\ref{matrixweight1}) using the generalized Schur decomposition \cite{ABu,GVL}
$X_j=U_j^\dag(Z_j+T_j)U_{j-1}$ for $j=1,\ldots,M$ with $U_j\in\U(N)$, $U_0=U_M$, $Z_j$ being complex diagonal $Z_j = {\mbox{diag}}(z_{j1},\ldots, z_{jN})$, and $T_j$ complex strictly upper triangular.
In this parametrization the matrix $X$ takes the form $X=U_M^\dag(Z+T)U_M$, where $Z=Z_M\cdots Z_1$ is diagonal and $T$ is upper triangular, i.e.
$Z+T = (Z_M+T_M)\cdots (Z_1+T_1) $. The diagonal elements of the $Z$-matrix play the role of the complex eigenvalues of $X$. In other words the eigenvalues of $X$ can be calculated as products of diagonal elements of the diagonal matrices $Z_j$ as $z_n = \prod_{j=1}^M z_{jn}$, for $n=1,\ldots,N$. Note that although the $z_n$'s are the eigenvalues of $X$, the $z_{jn}$'s are not the eigenvalues of $X_j$.  In this parametrization the measure $d\nu(X)$, see eq.~(\ref{matrixweight1}), yields
\begin{equation}
\begin{split}
\label{matrixweight2}
& d\nu(\{Z_i,T_i,U_i\}) \propto \left|\Delta\left(\prod_{k=1}^M Z_k\right)\right|^2
\\ & \times \prod_{i=1}^M\left(\int {\det}^{-L}[\imath H_i-\eins_N]\exp[{\tr (U_i^\dagger(Z_i+T_i)(Z_i+T_i)^\dagger U_i -\eins_N)(\imath H_i-\eins_N)}]\  d[H_i]\right)
d[Z_i] d[T_i] d \chi(U_i)  \ ,
\end{split}
\end{equation}
where
\begin{eqnarray}\label{Vand}
\Delta(Z)=\prod\limits_{1\leq a<b\leq N}(z_a-z_b)
\end{eqnarray}
is the Vandermonde determinant for $Z={\rm diag}(z_1,\ldots,z_N)$, $d\chi(U_i)$ is the Haar measure for the unitary group $\U(N)$, and $d[Z_i]$, $d[T_i]$ and $d[H_i]$ are flat measures for $Z_i$, $T_i$ and the $N\times N$ Hermitian matrix $H_i=H_i^\dagger$. The square of the Vandermonde determinant arises as a Jacobian for this parametrization \cite{ABu}.

The joint probability density function for the eigenvalues of the product matrix $X$ can be calculated by integrating out all degrees of freedom but $Z$ in the measure~(\ref{matrixweight2}),
\begin{equation}
 P^{(N,L,M)}(z_1,\ldots, z_N) = \int \delta(Z - Z_M\cdots Z_1) d\nu(\{Z_i,T_i,U_i\}).
\label{Pzzz}
\end{equation}
The integration over the $U_i$'s completely factorizes in eq.~\eqref{Pzzz} since they can be absorbed in the Hermitian matrices $H_i$'s. Then the integration over the $T_i$'s and $U_i$'s  (\ref{Pzzz}) reads
\begin{equation}
\label{jpdfZ}
 P^{(N,L,M)}(z_1,\ldots, z_N) \propto \int \delta(Z - Z_M\cdots Z_1) \left|\Delta\left(\prod_{k=1}^M Z_k\right)\right|^2 \prod_{i=1}^M W^{(L)}_1(Z_i) d[Z_i],
\end{equation}
where
\begin{equation}
W^{(L)}_1(Z_i) \propto \int {\det}^{-L}[\imath H_i-\eins_N]\exp[{\tr ((Z_i+T_i)(Z_i+T_i)^\dagger -\eins_N)(\imath H_i-\eins_N)}]\  d[H_i]d[T_i]
\label{W1}
\end{equation}
for $i=1,\ldots,M$. The integration over each $T_i$ and $H_i$ can be done explicitly \cite{KS},
\begin{equation}
W^{(L)}_1(Z_i) \propto \prod_{n=1}^N w^{(L)}_1(z_{in}) =
\prod_{n=1}^N \frac{L}{\pi}  (1-|z_{in}|^2)^{L-1} \Theta(1-|z_{in}|).
\label{Ww1}
\end{equation}
Note that this result is true for any $L>0$ independently of whether or not the condition $L\geq N$ is fulfilled. We briefly recall the  calculation of this integral in appendix~\ref{app2}. The constant $L/\pi$ is added for convenience in order to ensure the normalization $\int w_1^{(L)}(z) d^2z = 1$.  Inserting this result into (\ref{jpdfZ}) we obtain
\begin{eqnarray}\label{jpdfnn}
 P^{(N,L,M)}(z_1,\ldots, z_N) \propto \left|\Delta_N\left(Z\right)\right|^2 W^{(L)}_M(Z)\ ,
\end{eqnarray}
with a weight $W^{(L)}_M(Z)$ which factorizes into a product over one-point weights for the diagonal elements of $Z={\rm diag}(z_1,\ldots,z_n)$,
\begin{equation}
W^{(L)}_M(Z) = \prod_{n=1}^N w_M^{(L)}(z_n) \ .
\end{equation}
The individual one-point weights are given by
\begin{equation}\label{1weight.1}
w_M^{(L)}(z_n)= \int\limits_{\mathbb{C}^M} \delta^{(2)}\left(z_n-\prod\limits_{j=1}^M z_{jn}\right)\prod\limits_{i=1}^M w^{(L)}_1(z_{in}) d^2 z_{in}.
\end{equation}
The weight $w_M^{(L)}$ to be computed in the next subsection \ref{sec2.2} only depends on the modulus of the argument $w_M^{(L)}(z)=w_M^{(L)}(|z|)$. The one-point weights $w_M^{(L)}$ for the product of $M$ matrices are constructed from the one-point weights for the single matrices, see $w_1^{(L)}$ in eq.~(\ref{1weight.1}). Their definition is equivalent to the one of the probability density function for a random variable being the product of $M$ independently, identically distributed complex random numbers. The integral equation (\ref{1weight.1}) can be transformed into a factorized form via the Mellin transform as we shall see in the next subsection. The joint probability density function~(\ref{jpdfnn}) with the overall normalization factor can be written in the standard form \cite{Mehta}
\begin{equation}
P^{(N,L,M)}(z_1,\ldots, z_N) = \frac{1}{N! \prod_{j=0}^{N-1} h_j}
\prod_{n=1}^N w_M^{(L)}(z_n) \prod_{a<b}^N |z_b-z_a|^2
\label{jpdfev2}
\end{equation}
with
\begin{equation}
h_j = \int  w_M^{(L)}(z) |z|^{2j} d^2 z =
\left( \int w_1^{(L)}(z) |z|^{2j} d^2 z\right)^{M} = \binom{L+j}{L}^{-M} \ .
\label{moments}
\end{equation}
This normalization can be deduced by applying the orthogonal polynomial method in the complex plane \cite{BHJ,Mehta} to the weight $w_M^{(L)}(z)$. Since the weight only depends on the modulus of $z$, the corresponding orthogonal polynomials $p_j(z)$, defined through $\int w_M^{(L)}(z) p_i(z) p^*_j(z) d^2 z = h_i \delta_{ij}$, are monic, $p_j(z) = z^j$. The joint probability density function (\ref{jpdfev2}) can then be written in a determinantal form
\begin{equation}
\label{detform}
P^{(N,L,M)}(z_1,\ldots, z_N) = \frac{1}{N!} \det_{1\leq a,b \leq N}\left[ K^{(N,L,M)}(z_a,z_b)\right]\ ,
\end{equation}
with the kernel
\begin{equation}\label{kernel}
K^{(N,L,M)}(u,v) = \sqrt{w_M^{(L)}(u)w_M^{(L)}(v)} \; T^{(N-1,L,M)}(uv^*)
\end{equation}
which is given in terms of the following truncated series
\begin{equation}
T^{(N-1,L,M)}(x) = \sum\limits_{j=0}^{N-1} \binom{L+j}{j}^{M} x^j   \ .
\label{ts}
\end{equation}
The upper limit of the sum corresponds to the value of the first superscript of $T^{(N-1,L,M)}$, in this case $N-1$. The determinantal form (\ref{detform}) is particularly helpful when one wants to calculate the $k$-point correlation function \cite{Mehta}
\begin{equation}\label{kcorrel}
R_k^{(N,L,M)}(z_1,\ldots,z_k)\equiv
\frac{N!}{(N-k)!}\int\limits_{\mathbb{C}^{N-k}} \!\!P^{(N,L,M)}(z_1,\ldots,z_N) \;
d^2 z_{k+1} \ldots d^2 z_N
=\det_{1\leq a,b\leq k}\left[ K^{(N,L,M)}(z_a,z_b)\right] .
\end{equation}
The eigenvalue density can be readily read off from Eq. (\ref{kcorrel}) as
\begin{equation}\label{leveldens}
R_1^{(N,L,M)}(z) = K^{(N,L,M)}(z,z) =
w^{(L)}_M(|z|) T^{(N-1,L,M)} (|z|^2) .
\end{equation}
In Sec.~\ref{sec3} we study the asymptotic behavior of the kernel (\ref{kernel}) and of the eigenvalue density~(\ref{leveldens}) in various limits. To this end we need to independently calculate the asymptotic behavior of the one-point weight functions $w_M^{(L)}(z)$, eq.~\eqref{1weight.1}, and of the truncated series $T^{(N-1,L,M)}(z)$ in eq.~\eqref{ts}. In the ensuing subsections we discuss the one-point weight function and then the truncated series.

For the joint probability density function and the kernel in the more general case of truncations resulting from matrices of different sizes $N+L_j$ we refer the reader to appendix~\ref{sec2.3}.

\subsection{One-point weight functions}\label{sec2.2}

We will now determine the family of one-point weights~\eqref{1weight.1} $w_M^{(L)}(z)$ recursively, which only depend on the modulus $|z|$.
For $M=1$ the weight is given by $w_1^{(L)}(z)=(L/\pi) (1-|z|^2)^{L-1} \Theta(1-|z|)$, see eq.~\eqref{Ww1}. For $M\geq2$ we proceed as follows. Using polar coordinates $z_{i}=r_i \e^{\imath\varphi_i}$ one can represent the integral (\ref{1weight.1}) as an $M$-fold Mellin-convolution
\begin{eqnarray}\label{1weight.2b} w_M^{(L)}(z)&=&\frac{(2L)^M}{2\pi}\int\limits_{[0,1]^{M}}
\delta\left(|z|-\prod\limits_{m=1}^{M}r_m\right) \prod\limits_{i=1}^{M}\left(1-r_i^2\right)^{L-1}{dr_i} \ ,
\end{eqnarray}
which can be turned into the standard $M$-fold convolution by a further substitution, $r_i=\e^{-\vartheta_i/2}$. This changes the multiplicative constraint in the Dirac delta function to an additive one
\begin{eqnarray}\label{1weight.2} w_M^{(L)}(z)&=&\frac{L^M}{\pi}\int\limits_{\mathbb{R}_+^{M}}\delta\left(2{\rm ln}|z|+\sum_{m=1}^{M}\vartheta_m\right) \prod\limits_{i=1}^{M}\left(1-\e^{-\vartheta_i}\right)^{L-1}d\vartheta_i \ .
\end{eqnarray}
These expressions will be helpful in the following sections to derive the asymptotic behavior of the weight for large argument. Alternatively one can take advantage of the following recursion relations in order to determine the weight,
\begin{equation}\label{recursion}
w_{M+1}^{(L)}(z)
= 2\pi \int\limits_0^1w_1^{(L)}(r) w_M^{(L)}\left(\frac{z}{r}\right) \frac{dr}{r}
= 2\pi \int\limits_0^1w_1^{(L)}\left(\frac{z}{r}\right) w_M^{(L)}(r) \frac{dr}{r}\ .
\end{equation}
They directly follow from eq.~(\ref{1weight.1}). As we show below these relations are especially helpful when one wants to explicitly determine the form of weights for given $L$ and $M$ in terms of elementary functions. Note that for weights with the support $|z|\le 1$ the integrand in the last equation is non-zero only for $r \in [|z|,1]$, so one can, in this case, replace the lower integration limit by $|z|$.

Since the weight function $w_M^{(L)}(z)$ depends on its argument $z$ only through the modulus $|z|$, it is convenient to introduce an auxiliary function $\Omega_M^{(L)}(x)$  with a real positive argument
\begin{equation}\label{defomega}
w_{M}^{(L)}(z) \equiv \frac{1}{\pi} \Omega_M^{(L)}(|z|^2) \ .
\end{equation}
The support of $\Omega_{M}^{(L)}$ is $[0,1]$. The factor $1/\pi$  ensures its normalization,
\begin{equation}\label{norm}
\int_0^1 \Omega_M^{(L)}(x) dx  = \int_0^1
\Omega_M^{(L)}\left(|z|^2\right) d|z|^2 =\int_{|z|\le 1} w_M^{(L)}(z) d^2 z  = 1\ ,
\end{equation}
and keeps the recurrence relation simple,
\begin{equation}\label{recursion2}
\Omega_{M+1}^{(L)}(x)
= \int_0^1 \Omega_1^{(L)}(y) \Omega_{M}^{(L)}\left(\frac{x}{y}\right) \frac{dy}{y}
= \int_0^1 \Omega_{1}^{(L)}\left(\frac{x}{y}\right) \Omega_M^{(L)}(y) \frac{dy}{y}\ .
\end{equation}
The lower integration limit for this particular combination of weights can be replaced by $x$ since the
integrand is zero for $y<x$. The Mellin transform
\begin{equation}\label{Mellin}
{\rm M}_{M}(s) = \int_0^1 \Omega_M^{(L)}(x) x^{s-1} dx
\end{equation}
is the key to solve the recursion~\eqref{recursion2}. The recursion
(\ref{recursion2})  assumes the form
\begin{equation}\label{Mellinrecursion}
{\rm M}_{M+1}(s) = {\rm M}_M(s) {\rm M}_1(s) = ({\rm M}_1(s))^{M+1}.
\end{equation}
The simplicity of this relation uncovers the importance of the Mellin transform for product matrices. The inverse Mellin transform\footnote{The constant in the contour is chosen as $c>k+1$, such that the $k$-th moment of the weight is bounded.} yields the following one-point weight function
\begin{equation}
w_M^{(L)}(z) = \frac{1}{\pi} \Omega_M^{(L)}\left(|z|^2\right) =
\frac{1}{\pi} \frac{1}{2\pi i} \int_{c-i\infty}^{c+i\infty}
({\rm M}_1(s))^M |z|^{-2s} ds \ .
\label{IMT}
\end{equation}
It remains to calculate the Mellin transform of the weight function for $M=1$, $w_1^{(L)}(z)$. It reads
\begin{equation}\label{Mellintrafow1}
{\rm M}_1(s) =\pi\int_0^1 w_1^{(L)}(\sqrt{x}) x^{s-1} dx= L\int_0^1 (1-x)^{L-1} x^{s-1} dx =
\frac{\Gamma(s)\Gamma(L+1)}{\Gamma(s+L)} = \binom{s+L-1}{s-1}^{-1}.
\end{equation}
Employing Eq.~\eqref{IMT}, we see that the one-point weight is
\begin{eqnarray}
 w_M^{(L)}(z)&=&\frac{(L!)^M}{\pi}\int_{\mathcal{C}} \left(\frac{\Gamma(-u)}{\Gamma(L-u)}\right)^M|z|^{2u} \frac{du}{2\pi\imath}\ \Theta(1-|z|)\nonumber\\
 &=&\frac{(L!)^M}{\pi}G^{M,\,0}_{M,\,M}\left(\mbox{}_{0,\ldots,0}^{L,\ldots,L} \bigg| \,|z|^2\right)\Theta(1-|z|)\nonumber\\
 &=&\frac{(L!)^M}{\pi}|z|^{2L} G^{M,\,0}_{M,\,M}\left(\mbox{}_{-L,\ldots,-L}^{0,\ldots,0} \bigg| \,|z|^2\right)\Theta(1-|z|) \ .\label{1weight.4}
\end{eqnarray}
The expression involves a Meijer G-function, see \cite{Grad}, for which we have used an identity in the last line.  The contour $\mathcal{C}$ goes from $-\imath\infty$ to $+\imath\infty$ and leaves the poles of the Gamma functions in the numerator to the right of the contour line. For different truncations $N+L_j\to N$ of the single matrices in the product matrix this weight can be trivially generalized by replacing the prefactor $(L!)^M $ by $L_1!\cdots L_M!$ and the indices $L$ in the upper row of the Meijer G-function by the $L_j$'s, see appendix~\ref{sec2.3}.

For small $M$ and $L$ the formulae \eqref{1weight.4} can be expressed in terms of elementary functions. Let us first
illustrate this for $L=1$. In this case $\Omega_1^{(1)}(x)=1$ for $x\in [0,1]$
and $0$ otherwise. The function $\Omega_M^{(1)}(x)$ vanishes for $x \notin [0,1]$, too.
In particular we have
\begin{equation}
w_2^{(1)}(z) = \frac{1}{\pi}\int_0^1 \Omega_1^{(1)}(y) \Omega_1^{(1)}\left(\frac{|z|^2}{y}\right) \frac{dy}{y} = \frac{1}{\pi}\int_{|z|^2}^1 \frac{dy}{y}\ \Theta(1-|z|)=\frac{ {\rm ln} \left(|z|^{-2}\right)}{\pi}\ \Theta(1-|z|)\ ,
\end{equation}
and generally
\begin{equation}\label{omegaL1}
w_{M+1}^{(1)}(z) = \frac{1}{\pi}\int_{|z|^2}^1 \Omega_{M}^{(1)}\left(\frac{|z|^2}{y}\right) \frac{dy}y \Theta(1-|z|)=
\frac{{\rm ln}^M\!\left(|z|^{-2}\right)}{\pi M!}\ \Theta(1-|z|)\ ,
\end{equation}
as can be easily seen by induction.
Similarly one can find the consecutive $\Omega_M^{(2)}$'s for $L=2$
\begin{equation}\label{specialcases}
\begin{split}
\Omega_1^{(2)}(x) & =2( 1 - x)\ \Theta(1-x)\ , \\
\Omega_2^{(2)}(x) & = \Big(-8(1-x) -4 (1+x){\rm ln} (x)\Big) \Theta(1-x)\ , \\
\Omega_3^{(2)}(x) & = \Big(48(1 - x) +24 (1+x){\rm ln} (x) +4(1-x){\rm ln}^2 (x)\Big) \Theta(1-x)\ ,
\quad {\rm etc.}
\end{split}
\end{equation}
or do the same for larger $L$. The function $\Omega_M^{(L)}$ always takes the form of a polynomial of order $L-1$ in $x$ and of order $M-1$ in ${\rm ln} (x)$ because of the general relation
\begin{eqnarray}\label{generalomega}
\Omega_{M}^{(L)}(x)&=&(L!)^M\int_{\mathcal{C}} \frac{x^{u}}{\prod_{l=0}^{L-1}(l-u)^M} \frac{du}{2\pi\imath}\Theta(1-x)\\
&=& \frac{(L!)^M}{(M-1)!}\sum\limits_{l=0}^{L-1}\left.\frac{\partial^{M-1}}{\partial \mu^{M-1}}\left(\frac{x^{l-\mu}}{\prod_{\substack{1\leq j\leq L \\ j\neq l}}(j-l+\mu)^M}\right)\right|_{\mu=0}\Theta(1-x)\ ,\nonumber
\end{eqnarray}
where we have employed the residue theorem. In general the term $\Omega_{M+1}^{(L)}(x) \sim  {\rm ln}^{M} (1/x)$ dominates the region around the origin which is exactly the same as for $L=1$. For $L=1$ we obtain eq.~\eqref{omegaL1}, and for $L=2$ we have
\begin{eqnarray}\label{omegaL2}
 \Omega_{M}^{(2)}(x)&=&\frac{2^M}{(M-1)!}\left.\frac{\partial^{M-1}}{\partial \mu^{M-1}}\left(\frac{x^{-\mu}}{(\mu+1)^M}+\frac{x^{1-\mu}}{(\mu-1)^M}\right)\right|_{\mu=0}\Theta(1-x)\\
 &=&\frac{2^M}{(M-1)!}\sum\limits_{m=0}^{M-1}(-1)^{M+m}
\frac{(2M-2-m)!}{(M-1-m)!m!}\left(x+(-1)^{m+1}\right){\rm ln}^m (x)\ \Theta(1-x)\ ,\nonumber
\end{eqnarray}
agreeing with the special cases in eqs.~\eqref{specialcases}.

\subsection{Truncated series}

In this subsection we list some useful representations of the truncated series (\ref{ts}) which is simply the kernel (\ref{kernel}) at finite-$N$ without the weight functions,
\begin{eqnarray}
 T^{(N,L,M)}(x)&=&\sum\limits_{j=0}^{N} \binom{L+j}{j}^{M} x^{j}\label{partrep1}\\
 &=&\frac{1}{(L!)^M}\left. \left(\prod\limits_{j=1}^M\frac{\partial^L}{\partial y_j^L}\right) \left(\frac{1-(y_1\cdots y_M)^{N+L+1}}{1-y_1\cdots y_M}\right)\right|_{y_j=x^{1/M}}\label{partrep2}\\
 &=&\int_0^{2\pi}\frac{d\varphi}{2\pi}\, _MF_{M-1}\left(\mbox{}_{\quad  1,\ldots,1}^{L+1,\ldots,L+1} \bigg| \,\e^{-\imath \varphi}\right)\frac{1-(x\e^{\imath\varphi})^{(N+1)}}{1-x\e^{\imath \varphi}}\label{partrep3}\\
 &=&N\int\limits_{0}^1\binom{L+N\xi}{N\xi}^{M} x^{N\xi}\sum\limits_{j=0}^{N} \delta(j-N\xi)d\xi.\label{partrep4}
\end{eqnarray}
Each representation will become useful when deriving specific asymptotic expansions.

Note that apart from the representation~\eqref{partrep2} all representations of the truncated series $T^{(N,L,M)}(x)$ can be easily generalized to a product of matrices with different truncations as it is discussed in appendix~\ref{sec2.3}. In the representations~\eqref{partrep1} and \eqref{partrep4} one has only to replace the exponentiated binomials by a product of binomials with different $L_j$'s and in the representation~\eqref{partrep3} one has to replace the indices $L+1$ in the upper row of the hypergeometric function $\ _MF_{M-1}$ by $L_j+1$. This generalization is not that simple for the representation~\eqref{partrep2} as can be easily checked by writing the rational function as a geometric sum.

\sect{Asymptotic behavior for large \boldmath $N$}\label{sec3}

In this section we consider the asymptotic behavior of the model for large $N\rightarrow \infty$ and for fixed $M$. Other important limits are postponed to forthcoming publications. We distinguish two cases: (1) an extensive truncation where the number of truncated columns and rows, $L$, is of order $N$, and (2) a weak truncation where $L$ is constant. The former one corresponds to an ensemble of matrices that strongly breaks the unitarity of the matrices drawn from the original  ensemble while the latter one corresponds to an ensemble of weakly non-unitary matrices.

\subsection{Strong non-unitarity: \boldmath $N,L \rightarrow \infty$ and $\alpha=L/N$ fixed}\label{sec3.2}

Let us denote
the ratio of the size $N$ of the truncated matrices and the size $N+L$ of the original unitary matrices before truncation by
\begin{equation}\label{defmualp}
\mu \equiv \frac{N}{N+L} = \frac{1}{1+\alpha}<1 \ \ \mbox{and}\ \ 0<\alpha\equiv \frac{L}{N}.
\end{equation}
We keep $\mu<1$ fixed while taking the limit $N\rightarrow \infty$.

First we study the macroscopic behavior in subsection~\ref{sec3.2.2} and derive the limiting eigenvalue density. In particular we show that the limiting eigenvalue density covers a disk centered at the origin of the complex plane with radius $\mu^{M/2}<1$. This is very different from the eigenvalues of a unitary matrix distributed on the unit circle. We call this limit with such an extensive truncation, $L\sim N$, the strong non-unitarity regime.

In subsections~\ref{sec3.2.2a} and~\ref{sec3.2.2b} we turn to the microscopic limit, showing that the eigenvalue fluctuations in the bulk and at the edge of the support of the macroscopic eigenvalue density have the same local universal behavior as the Ginibre ensemble. Finally in subsection~~\ref{sec3.2.1} we discuss the behavior of the kernel at the origin
and show that it falls into the same universal classes labeled by $M$ as the product of Ginibre matrices \cite{ABu}.


\subsubsection{Macroscopic regime}\label{sec3.2.2}

Here we calculate the mean limiting eigenvalue density for $N\rightarrow \infty$.
We first derive the asymptotic behavior of the truncated series~\eqref{ts} and, then, of the one-point weights~(\ref{1weight.4}) in the large $N$ limit while keeping the eigenvalues of order unity, $0<|z|<1$. These two results constitute the eigenvalue density~\eqref{leveldens}.

For the truncated series we employ the representation~\eqref{partrep4}. The sum of Dirac delta-functions can be omitted to leading order in the $1/N$-expansion. This amounts to substituting the sum over $j$ by an integral over $\xi=j/N$ running from zero to one and normalized to unity. The binomial symbol containing $\xi$ in the remaining integral is approximated by Stirling's formula, i.e.
\begin{eqnarray}\label{bino}
 \binom{L+N\xi}{N\xi} \approx \sqrt{\frac{\alpha+\xi}{2\pi \xi L}}\ \alpha^{-L}\exp\left[N\left(\left(\alpha+\xi\right){\rm ln}\left(\alpha+\xi\right)-\xi{\rm ln}\xi\right)\right]\ ,
\end{eqnarray}
so we have (\ref{partrep4})
 \begin{equation}
 \label{partmacro1}
 T^{(N,L,M)}(|z|^2)\approx  \sqrt{\frac{N^2}{(2\pi L)^M}}\ \alpha^{-ML} \int\limits_{0}^1\left(\frac{\alpha+\xi}{\xi}\right)^{M/2}
\exp\left[Nf(\xi)\right]d\xi
\end{equation}
with the real function
\begin{eqnarray}\label{actionf}
 f(\xi)=M\left(\left(\alpha+\xi\right){\rm ln}\left(\alpha+\xi\right)-\xi{\rm ln}\xi+\xi {\rm ln}|z|^{2/M}\right).
\end{eqnarray}
The large $N$ behavior of this integral can be found by employing Laplace' method (saddle point method for a real function in the exponent).
The idea is to replace the exponential function of the integrand by a Gaussian with the width of order $1/\sqrt{N}$ and the maximum located at $\xi_0$ given by the saddle point equation
 \begin{equation}\label{saddleeq}
 \ln\left(\frac{\alpha+\xi_0}{\xi_0}\right)+ \ln\left(|z|^{2/M}\right)=0 \ .
\end{equation}
This yields for the saddle point
\begin{equation}\label{saddle pointxi}
\xi_0 =  \frac{\alpha}{|z|^{-2/M}-1} = \frac{\mu^{-1}-1}{|z|^{-2/M}-1} \ .
\end{equation}
In case of varying truncations, $L_j\neq L_i$ for $j\neq i$, (see appendix~\ref{sec2.3}) the corresponding saddle point equation would be more involved.  The saddle point equation~\eqref{saddleeq} is generically equivalent to a polynomial equation of order $M$.

For $|z|<\mu^{M/2}$ the maximum of $f(\xi)$ is located inside the integration range $[0,1]$ of the integral (\ref{partmacro1}). When $|z|$ is close to $\mu^{M/2}$, or more precisely when $|z| - \mu^{M/2}$ is of order $1/\sqrt{N}$, a part of the Gaussian lies inside the integration range and a considerable part outside, while when $|z| - \mu^{M/2} \ll 1/\sqrt{N}$ the whole Gaussian lies outside the integration range. This means that in this approximation the integral~(\ref{partmacro1}) is given by the integral over the Gaussian times a factor corresponding to the fraction of the Gaussian that lies inside the integration range. This factor is approximated by a complementary error function,  $\erfc\left(a \sqrt{N} (|z|-\mu^{M/2})\right)$, with some positive constant $a$. This function changes its value from one to zero for $|z|$ in a narrow interval whose width is of order $1/\sqrt{N}$. The width tends to zero for $N\rightarrow \infty$ and the function reduces to the Heaviside step function
\begin{eqnarray}\label{partmacro1b}
T^{(N,L,M)}(|z|^2)&\approx&\sqrt{\frac{1}{M(2\pi L)^{M-1}}}|z|^{(1-M)/M}\left(1-|z|^{2/M}\right)^{-LM-1}\Theta(\mu^{M/2}-|z|) .
\end{eqnarray}
One should remember that for large but finite $N$ the Heaviside step function in the last equation should be rather replaced by the complementary error function. We will come back to this point in subsection~\ref{sec3.2.2b} when discussing the fluctuations in the microscopic limit at the edge of the spectrum.

To complete the derivation of the large $N$ limit for the kernel (\ref{kernel}) we also need to calculate the limiting weight $w^{(L)}_M(z)$, eq.  (\ref{1weight.1}). We determine it using eq. (\ref{1weight.2}). By integrating out $\vartheta_M$ we obtain an integral representation
\begin{equation}\label{weightmacro2a}
 w_M^{(L)}(z) = \frac{L^M}{\pi}\Theta(1-|z|)\int\limits_{\mathbb{R}_+^{M-1}} \left(1-|z|^2\exp\left[\sum_{m=1}^{M-1}\vartheta_m\right]\right)^{L-1}
 \prod\limits_{i=1}^{M-1}\left(1-\e^{-\vartheta_i}\right)^{L-1}d\vartheta_i\ ,
\end{equation}
which is well suited for a saddle point analysis for large $L$ and fixed $|z|$.
The maximum of the $(M-1)$-dimensional integrand is located at the point
$(\vartheta^{(0)}_1,\ldots,\vartheta^{(0)}_{M-1})$ given by the following $M-1$ equations:
\begin{eqnarray}\label{saddle1}
\frac{1}{1-\e^{\vartheta_j^{(0)}}}=\frac{|z|^2}{|z|^2-\exp[-\sum_{m=1}^{M-1}\vartheta_m^{(0)}]},\quad\forall j\in{1,\ldots, M-1}\ .
\end{eqnarray}
They have a unique symmetric solution $\vartheta_1^{(0)}=\ldots=\vartheta_{M-1}^{(0)}=-(2 {\rm ln}|z|)/M$. When expanding around this saddle point we have to calculate the determinant of the $(M-1)\times(M-1)$ Hessian matrix at the saddle point, i.e.
\begin{eqnarray}\label{detmat}
 \det\left[\frac{|z|^{-2/M}}{(1-|z|^{-2/M})^2}(1+\delta_{ab})\right]_{1\leq a,b\leq M-1}=\frac{M|z|^{2(M-1)/M}}{(1-|z|^{2/M})^{2(M-1)}}.
\end{eqnarray}
Summarizing the steps we find the asymptotic formula for $N\rightarrow \infty$
\begin{eqnarray}\label{weightmacro2b}
 w_M^{(L)}(z)& \approx &\sqrt{\frac{(2\pi L)^{M-1}}{M}}\frac{L}{\pi}|z|^{(1-M)/M}\left(1-|z|^{2/M}\right)^{ML-1}\Theta(1-|z|)
\end{eqnarray}
for the one-point weights. One should note that the corresponding expression for the one-point weights 
would be more complicated for varying truncations $L_j$ since in that case 
the saddle point equation for the counterpart of the integral~\eqref{weightmacro2a}
would have generically a non-symmetric solution. The symmetric solution is particular and does not apply to the general case of different truncations proposed in appendix~\ref{sec2.3}.

Inserting the eq.~(\ref{partmacro1b}) and eq.~\eqref{weightmacro2b} in eq.~(\ref{leveldens}) we obtain the normalized level density
\begin{equation}\label{densitymacro2}
 \rho^{(\mu,M)}(z)  \equiv \lim_{N,L\to\infty}\frac1N R_1^{(N,L,M)}(z)=
 \frac{\alpha}{\pi M}\frac{|z|^{2(1-M)/M}}{(1-|z|^{2/M})^2}\Theta(\mu^{M/2}-|z|).
\end{equation}
It agrees with the result derived in Ref.~\cite{KS,EB} for $M=1$ and \cite{BNS} for general $M$. We see that the support of this function is  a disk of radius $\mu^{M/2}$ which is smaller than the radius $\sqrt{\mu}$ of a single matrix ($M=1$) in the product since $\mu \le 1$. Note that by changing variables,
$z = |z| \e^{\imath \phi} \rightarrow v = |z|^{1/M} \e^{\imath \phi}$, one obtains
\begin{equation}
d^2 z \rho^{(\mu,M)}(z) = d^2 v \rho^{(\mu,1)}(v)\ ,
\end{equation}
which means that the product of $M$ independent truncated matrices has the same limiting density as the $M$-th power of a single random matrix drawn from the same ensemble for $N\rightarrow \infty$. This is due to the self-averaging property of isotropic matrices \cite{BNS}. This also agrees with what was found for the product of $M$ Ginibre matrices \cite{Ralf,ABu,IK,BJW,goetze}.
One should also note that the density can be made globally flat on its support by the following reparameterization $z \rightarrow \widehat{z}=\sqrt{(1-\mu)/(1-|z|^{2/M})} \e^{\imath\phi}$ that gives
\begin{equation}\label{densitymacro2b} \widehat{\rho}^{(\mu,M)}(\widehat{z})=
\left\{ \begin{array}{ll}
\displaystyle\frac{1}{\pi \mu} \ ,  &  \sqrt{1-\mu} \le |\widehat{z}| \le 1, \\
0 \ , & {\rm otherwise \ . } \end{array}
\right.
\end{equation}
This is the well-known limiting ring distribution of the Ginibre ensemble of rectangular matrices of dimensions $N\times (N+L)$ whose ratio $\mu=N/(N+L)$ is kept constant in the large $N$ limit, see ref.~\cite{Jonit}.

We conclude this subsection by considering truncations much larger than the remaining matrix, $L\gg N$.
Here the parameter $\mu$ becomes very small, being approximately equal to
$\mu \approx N/L \ll 1$. In this case the radius of the support of the eigenvalue density of the product $X^{(M)}$ is of order $(N/L)^{M/2}\ll1$. This can be fixed by rescaling all matrices in the product by a factor $(N/L)^{1/2}$, $X_j = (N/L)^{1/2} Y_j$,  and we obtain a product matrix $Y\equiv Y^{(M)} = Y_M \cdots Y_1$ whose eigenvalue density has a support of radius one. Moreover the probability measure for individual matrices $Y_j$ in the product becomes Gaussian in this limit, as follows from eq.~(\ref{measurej})
\begin{equation}
\begin{split}
d\mu\left(\sqrt{\frac{N}{L}} Y\right) &\propto  {\det}^{L-N}\left(\eins_{N}-\frac{N}{L}Y^\dagger Y \right)\Theta\left(\eins_{N}-\frac{N}{L} Y^\dagger Y \right)d[Y] \\
&\approx \exp\left[-N {\rm Tr} Y^{\dagger} Y\right] d[Y] \ .
\end{split}
\end{equation}
In the second step of the calculation we assumed that $N/L \ll 1$ and used the following approximation $\det(\eins - \alpha^{-1} A) = \exp({\rm Tr} \ln (\eins - \alpha^{-1} A)) \approx \exp( -\alpha^{-1} {\rm Tr} A)$ that holds for $|\alpha^{-1}| \ll 1$.
We also omitted the Heaviside function because the matrix in its argument becomes automatically positive definite in the limit $\alpha^{-1}=N/L \rightarrow 0$.
In other words in the regime $L\gg N$ the behavior of the product of independent truncated unitary matrices is equivalent to the product of independent Ginibre matrices.

\subsubsection{Bulk fluctuations}\label{sec3.2.2a}

In the previous subsection we determined the large $N$ asymptotic behavior of the truncated series~(\ref{partmacro1b}),
\begin{eqnarray}\label{3series}
T^{(N,L,M)}(uv^*)&\approx&{\frac{1}{\sqrt{M(2\pi L)^{M-1}}}}(uv^*)^{(1-M)/(2M)}\left(1-(uv^*)^{1/M}\right)^{-LM-1}\Theta(\mu^{M/2}-|z|) \ ,  \ \
\end{eqnarray}
and of the one-point weight (\ref{weightmacro2b}),
\begin{eqnarray}\label{3weight}
 w_M^{(L)}(u)& \approx &\sqrt{\frac{(2\pi L)^{M-1}}{M}}\frac{L}{\pi}|u|^{(1-M)/M}\left(1-|u|^{2/M}\right)^{ML-1}\Theta(1-|z|)\ ,
\end{eqnarray}
which we repeat here for completeness. Together they build up the kernel~(\ref{kernel}) which is given by
\be
K^{(N,L,M)}(u,v) = \sqrt{w_M^{(L)}(u)w_M^{(L)}(v)} \; T^{(N-1,L,M)}(uv^*)\ .
\label{kernel_muM}
\ee
Note that with complex arguments the truncated series contains a phase, whereas the weights only depending on the modulus do not.
These formulae hold inside the disk of radius $\mu^{M/2}$. The kernel vanishes when either $u$ or $v$ lie outside the disk.

In this subsection we determine the microscopic large $N$ limit in the bulk of the spectrum. This can be analyzed by studying the kernel (\ref{kernel_muM}) for two neighboring points,
\begin{equation}
u = z + \frac{1}{\sqrt{N}} \delta u \quad {\rm and} \quad
v= z + \frac{1}{\sqrt{N}}\delta v,
\label{bulkpoint}
\end{equation}
both located in the vicinity of the point $z$ in the bulk, i.e.  $0< |z| < \mu^{M/2}$, with $|z|$, $|\delta u|$ and $|\delta v|$ of order unity.
The calculation is carried out as a $1/\sqrt{N}$-expansion up to second order, by inserting (\ref{bulkpoint}) into the expressions for the limiting truncated series~\eqref{3series} and weights~\eqref{3weight}. Technically it is easier when the $u$- and $v$-dependent factors are first exponentiated and then Taylor expanded in the two variables in the standard way,
\bea
\exp[g(u,v^*)]&\approx&
\exp\left[ g(z,z^*)+\frac{1}{\sqrt{N}}(\delta u,\delta v^*)(\partial_u g,\partial_{v^*}g)^T\right.\nn\\
&&\left.+\frac{1}{2N}
(\delta u,\delta v^*)
\left(\begin{array}{cc}
\partial_{u}\partial_{u} g& \partial_{u}\partial_{v^*}g\\
\partial_{v^*}\partial_{u} g& \partial_{v^*}\partial_{v^*} g
\end{array}
\right)
(\delta u,\delta v^*)^T
\right]
\eea
up to higher order terms, where $g$ is a general action. In the limit of large $N$ and $L$ only the factors with an exponent proportional to $L$ will contribute. We obtain the following result for the truncated series
 \begin{eqnarray}
T^{(N,L,M)}(uv^*)&\approx&{\frac{1}{\sqrt{M(2\pi L)^{M-1}}}}\ |z|^{(1-M)/(M)}\left(1-|z|^{2/M}\right)^{-LM-1}
\nonumber\\
&&\times
\exp\left[\frac{L|z|^{2(1-M)/M}}{\sqrt{N}(1-|z|^{2/M})}(\delta u z^*+\delta v^* z)
+\frac{L|z|^{2(1-M)/M}}{NM(1-|z|^{2/M})^2}\delta u \delta v^*
\right]\nonumber\\
&&\times
\exp\left[-\frac{L|z|^{2(1-2M)/M}}{2N(1-|z|^{2/M})}\left(1
-\frac{1}{M(1-|z|^{2/M})}\right)(\delta u^2z^{*\,2}+\delta v^{*\,2}z^2)\right],\ \
\label{partbulk2}
\end{eqnarray}
and for the weight at argument $u$
\begin{eqnarray}
 w_M^{(L)}(u)&\approx&\sqrt{\frac{(2\pi L)^{M-1}}{M}}\frac{L}{\pi}|z|^{(1-M)/M}\left(1-|z|^{2/M}\right)^{ML-1}\nonumber\\
 &&\times\exp\left[-\frac{L|z|^{2(1-M)/M}}{\sqrt{N}(1-|z|^{2/M})}(\delta u z^*+\delta u^* z)
 - \frac{L|z|^{2(1-M)/M}}{NM(1-|z|^{2/M})^2}\delta u\delta u^*\right]\nonumber\\
 &&\times\exp\left[\frac{L|z|^{2(1-2M)/M}}{2N(1-|z|^{2/M})}\left(1-\frac{1}{M(1-|z|^{2/M})}\right)
(\delta u^2z^{*\,2}+\delta u^{*\,2}z^2)
\right].\label{weightbulk2}
\end{eqnarray}
Inserting these two asymptotic results into eq.~(\ref{kernel_muM}) we obtain the following expression for the limiting microscopic kernel in the bulk of the spectrum,
\bea
 K^{(\mu,M)}_{\rm bulk}\left(\delta u,\delta v\right) &\equiv& \lim_{N,L\to\infty}\frac{1}{N}
K^{(N,L,M)}(u,v)\nonumber\\
&=&
\rho^{(\mu,M)}(z) \e^{\imath \Phi(\delta u,\delta v)}
\exp\left[ -2 \pi \rho^{(\mu,M)}(z)\left(\frac{1}{2}|\delta u|^2 + \frac{1}{2} |\delta v|^2 - \delta u \delta v^*\right)\right]
\label{kerneldudvbulk}
\eea
where we have introduced the phase
\bea
\Phi(\delta u,\delta v)&\equiv& \frac{L|z|^{2(1-M)/M}}{\sqrt{N}(1-|z|^{2/M})}\Im m((\delta u-\delta v) z^*)
\nn\\
&&-\frac{L|z|^{2(1-2M)/M}}{2N(1-|z|^{2/M})}\left(1-\frac{1}{M(1-|z|^{2/M})}\right)
\Im m\Big((\delta u^{2}-\delta v^{2})z^{*\,2}\Big)\ .
\label{phase-bulk}
\eea
It is a real function which is antisymmetric $\Phi(\delta u,\delta v) = - \Phi(\delta v,\delta u)$ under exchanging its arguments. Thus the phase factor cancels out in the expression for the $k$-point  correlation functions given by the determinants of the kernel (\ref{kcorrel}). The remaining part of the expression~\eqref{kerneldudvbulk} contains the macroscopic density (\ref{densitymacro2}) at the point $z$ where we zoom in. It is equivalent to the Ginibre kernel \cite{Ginibre} which is universal \cite{ameur,berman,TV}. When changing variables to
\be
\delta \hat{u} = \sqrt{2 \pi \rho^{(\mu,M)}(z)}\ \delta {u}  \quad , \quad
\delta \hat{v} = \sqrt{2 \pi \rho^{(\mu,M)}(z)}\ \delta {v}
\ee
we obtain the identical expression for the microscopic bulk kernel as in the Ginibre ensemble \cite{Ginibre}.

As a last remark, note that in the previous subsection~\ref{sec3.2.2} on the macroscopic behavior we found in the limit $\mu\to0$ the product of Ginibre matrices. In this limit, the universal microscopic bulk kernel of a product of Ginibre matrices was already found \cite{ABu}. Additionally universality in the bulk holds for all values of $1>\mu\geq0$ including the limit $\mu\to0$.

\subsubsection{Edge fluctuations}\label{sec3.2.2b}

We are going to derive the asymptotic behavior of the kernel (\ref{kernel}) in the large $N$ limit for fixed $\alpha=N/L$, for $u,v$ lying close to each other and close to the edge. More precisely, we consider
\be
u = \mu^{M/2}\e^{\imath\varphi} + \frac{1}{\sqrt{N}} \delta u\ \ \mbox{and}\ \  v= \mu^{M/2}\e^{\imath\varphi} + \frac{1}{\sqrt{N}}\delta v
\label{edge-scaling}
\ee
in the vicinity of a point $z=\mu^{M/2}\e^{\imath\varphi}$  on the edge of the support of the distribution \eqref{densitymacro2}, and $|\delta u|$ and $|\delta v|$ are both of order unity. Since the level density is rotationally invariant we assume that $\varphi=0$ without loss of generality.

As before we begin with calculating the truncated  series~\eqref{partmacro1} using the saddle point method, except that now we do this for the sum $T^{(N,L,M)}(uv^*)$ of a complex argument $uv^*$. We obtain an analogous expression as~\eqref{partmacro1} except that
$\ln |z|^2$ is replaced by the principal value of $\ln (uv^*)$
\begin{equation}\label{logexp}
\ln (uv^*) \approx M \ln (\mu) +\frac{\mu^{-M/2}}{\sqrt{N}}(\delta u+\delta v^*)-\frac{\mu^{-M}}{2N}(\delta u^2+\delta v^{*\,2})\equiv  M\ln(\mu)+A+B\ .
\end{equation}
Here we define the terms $A$ and $B$ of order $1/\sqrt{N}$ and $1/N$, respectively, while neglecting higher order terms in the expansion. The saddle point
$\xi_0$ fulfils an equation analogous to (\ref{saddleeq}), except that $|z|^{-2/M}$ is replaced by $(uv^*)^{-1/M}$, and is expanded as in eq.~(\ref{logexp}),
\be
0=f'(\xi_0)= M\ln\left( \frac{\alpha+\xi_0}{\xi_0}\right)+M\ln(\mu)+A+B.
\label{fprime}
\ee
Recall the definition of the action $f$ in eq.~\eqref{actionf}.
This leads to the following expression for the solution $\xi_0$ of the saddle point equation, expanded up to leading order:
\be
\xi_0\approx1 + \frac{\mu^{-M/2}}{\sqrt{N}\ M(1-\mu)}(\delta u+\delta v^*)\equiv 1+\frac{A}{M(1-\mu)}.
\label{sp-edge}
\ee
This expansion can be also directly obtained from eq.~\eqref{saddle pointxi}. The saddle point $|\xi_0|\leq 1$ is located in the vicinity of the upper integration interval at $\xi=1$. Therefore we can no longer replace the integral by a Gaussian integration over the whole real axis when expanding $\xi=\xi_0-\delta\xi/\sqrt{N}$. Instead we have to set a lower bound of the integral over $\delta\xi$ at $A/(M(1-\mu))$. Integrating over $\delta\xi$ we obtain a complementary error function,
\be
T^{(N,L,M)}(u,v^*)\approx  \sqrt{\frac{N^2}{(2\pi L)^M}}\ \alpha^{-ML}
\exp[Nf(\xi_0)]\sqrt{\frac{\pi}{-2Nf''(\xi_0)}}\ \mbox{erfc}\left[\frac{A}{M(1-\mu)}\sqrt{-\frac{N f''(\xi_0)}{2}}\right]\ ,
\ee
given in terms of the notation from eqs.~(\ref{fprime}) and (\ref{sp-edge}) above, cf. eq.~\eqref{partmacro1}. While in the last two terms the limiting value $-f''(\xi_0=1)=M(1-\mu)$ is legitimate, we still need to expand the first exponential factor $\exp[Nf(\xi_0)]$ up to second order in $1/\sqrt{N}$ using eqs.~(\ref{logexp}) and (\ref{sp-edge}). Collecting all terms to that order we finally arrive at
\bea
\label{partedge2}
 T^{(N,L,M)}(uv^*) & \approx& \frac{(1-\mu)^{-LM-1}\mu^{-(M-1)/2}}{(2\pi L)^{(M-1)/2}\sqrt{M}}\exp\left[\sqrt{N}\mu^{-M/2}(\delta u+\delta v^*)\right]\\
 & \times&\exp\left[-\frac{\mu^{-M}}{2}(\delta u^2+\delta v^{*\,2})+\frac{\mu^{-M}}{2M(1-\mu)}(\delta u+\delta v^*)^2\right]{\rm erfc}\left[\frac{\mu^{-M/2}}{\sqrt{2M(1-\mu)}}(\delta u+\delta v^*)\right].\nn
\eea

Let us come to the calculation of the asymptotic behavior of  the one-point weight for large $N$ and $L$.
Applying the result~\eqref{weightmacro2b} to $u$ and $v$ in our edge scaling regime, see eq.~(\ref{edge-scaling}), the two leading terms of the $1/\sqrt{N}$ expansion yield
\begin{equation}\label{weightedge2}
\begin{split}
 w_M^{(L)}(u)&\approx \sqrt{\frac{(2\pi L)^{M-1}}{M}}\frac{L}{\pi}\mu^{-(M-1)/2}\left(1-\mu\right)^{ML-1}
 \exp\left[-\sqrt{N}\mu^{-M/2}(\delta u +\delta u^*)\right]\\
 &\times\exp\left[\frac{\mu^{-M}}{2}\left(1-\frac{1}{M(1-\mu)}\right)(\delta u^2+\delta u^{*\,2})
 -\frac{\mu^{-M}}{M(1-\mu)}\delta u\delta u^* \right]
\end{split}
\end{equation}
and analogously for $w_M^{(L)}(v)$. Summarizing the results~(\ref{weightedge2}) and (\ref{partedge2}) the microscopic limit of the kernel at the edge is
\bea
\label{kerneledge2}
K^{(\mu,M)}_{\rm edge}\left(\delta u,\delta v\right) &\equiv& \lim_{N,L\to\infty}\frac{1}{N}
 K^{(N,L,M)}(u,v)\nn\\
 &=&\frac{\mu^{-M}}{\pi M(1-\mu)}\exp \left[ \imath \Phi(\delta u,\delta v) \right]
 \exp\left[-\frac{\mu^{-M}}{2M(1-\mu)}\left({|\delta u|^2+|\delta v|^2}-2\delta u\delta v^*\right)\right]
 \nn\\
 &\times&{\rm erfc}\left[\frac{\mu^{-M/2}}{\sqrt{2M(1-\mu)}}(\delta u+\delta v^*)\right] ,
\eea
where $\Phi(\delta u,\delta v) = -\Phi(\delta v,\delta u)$ is a real phase,
\begin{equation}
\Phi(\delta u, \delta v)= \sqrt{N}\mu^{-M/2}\Im m(\delta u -\delta v)+\frac{\mu^{-M}(1-M+\mu M)}{2M(1-\mu)}\Im m(\delta u^2-\delta v^2)\ .
\label{phi_kernel}
\end{equation}
The phase factors $\exp[\imath \Phi]$ cancel out in the determinantal structure of the $k$-point correlation function~\eqref{kcorrel} and, thus, are irrelevant in the spectral statistics. What remains is the universal error function kernel that agrees with that of the Ginibre ensemble \cite{Mehta,FH,EK03} after changing to new variables 
\be
\delta \hat{u}\equiv \frac{\mu^{-M/2}}{\sqrt{M(1-\mu)}}\delta u .
\ee
In particular the microscopic density at the edge is then given by
\begin{eqnarray}\label{densityedge2}
\rho^{(\mu,M)}_{\rm edge}(\delta u)
&=&
K^{(\mu,M)}_{\rm edge}\left(\delta u,\delta u\right)=
\frac{\mu^{-M}}{\pi M(1-\mu)}{\rm erfc}\left[\sqrt{\frac{2\mu^{-M}}{M(1-\mu)}}|\delta u|\right]\ ,
\end{eqnarray}
which is the universal result \cite{Mehta,FH,EK03}.

We expect that the universal results~\eqref{kerneledge2} and \eqref{densityedge2} also hold in the general setting with different truncations, see appendix~\ref{sec2.3}. Since the saddle point equations are highly complicated in this case we have not proven it here.

\subsubsection{Fluctuations at the origin}\label{sec3.2.1}

While studying the asymptotic behavior of the kernel in the vicinity of  the origin for
$N\rightarrow \infty$ it is convenient to  introduce a rescaled variable $\delta z$ with $|\delta z|$ of order one,
\begin{equation}
z = L^{-M/2} \delta z\ ,
\label{Ldz}
\end{equation}
and express results in $\delta z$. One could alternatively use a variable $\delta z'$
in the scaling formula $z = N^{-M/2} \delta z'$ but since  $\alpha=L/N$ is kept fixed in the limit $N\rightarrow \infty$, $\delta z$ and $\delta z'$ differ by an inessential constant $\delta z = \alpha^{M/2} \delta z'$ which does not affect the $N$-dependence of the scaling.

For $N\rightarrow \infty$ the truncated series asymptotically behaves like
\begin{equation}
\label{partmicro2}
T^{(N,L,M)}(|z|^2) \approx \sum\limits_{j=0}^{\infty} \frac{|\delta z|^{2j}}{[j!]^{M}}=\, _0F_{M-1}\left(\mbox{}_{1,\ldots,1}^{-} \bigg| \,|\delta z |^2\right)
\end{equation}
resulting from~\eqref{partrep1}. Note that this asymptotic result does not change at all when choosing different truncations of the matrices in the product matrix $X^{(M)}$, see appendix~\ref{sec2.3}.

The asymptotic behavior of the one-point weights (\ref{1weight.1}) can be derived from the first line of the integral representation~\eqref{1weight.4} using the following asymptotic behavior of the Gamma function:
\be
\lim_{L\to\infty}\frac{\Gamma(L-u)}{\Gamma(L)}\ \e^{u\ln(L)}=1\ .
\ee
This directly leads to
\begin{equation}
w_M^{(L)}(z)
\approx
\frac{L^M}{\pi}G^{M,\,0}_{0,\,M}\left(\mbox{}_{0,\ldots,0}^{-} \bigg| \,  |\delta z|^2\right)\ .
\label{weightmicro2c}
\end{equation}
Also this result can be readily generalized to the situation of different truncations $N+L_j\to N$ of the matrices in the product matrix, see appendix~\ref{sec2.3}, by replacing the prefactor $L^M$ by $L_1\cdots L_M$.

Now we have all constituents needed for the microscopic limit of the kernel~(\ref{kernel}) at the origin. Combining eqs.~(\ref{partmicro2}) and (\ref{weightmicro2c}) we obtain
\bea
K^{(\mu,M)}_{\rm origin}(\delta u,\delta v) &\equiv&
\lim_{N,L\to\infty} L^{-M} K^{(N,L,M)}(u+L^{-M/2}\delta u,v+L^{-M/2}\delta v)
\nn\\
&=& \frac{1}{\pi}\sqrt{G^{M,\,0}_{0,\,M}\left(\mbox{}_{0,\ldots,0}^{-} \bigg| \,|\delta u|^2\right)G^{M,\,0}_{0,\,M}\left(\mbox{}_{0,\ldots,0}^{-} \bigg| \,|\delta v|^2\right)}\ _0F_{M-1}\left(\mbox{}_{1,\ldots,1}^{-} \bigg| \,\delta u {\delta v}^*\right).\label{kernelmicro2a}
\eea
The kernel agrees with the microscopic kernel at the origin for the complex eigenvalues of products of Ginibre matrices  \cite{ABu} hinting to a universal property of product matrices. The same result follows for the model with different truncations described in appendix \ref{sec2.3}, whenever all $L_j\to\infty$. Only the rescaling has to be changed from $L^{-M/2}$ to $\prod_{j=1}^M L_j^{-1/2}$.

As a check, one can regain the Ginibre kernel \cite{Ginibre} from eq.~(\ref{kernelmicro2a}) for $M=1$,
\begin{equation}
K^{(\mu,M=1)}(u,v)=\frac{1}{\pi}
\exp\left[-\frac{|\delta u|^2}{2}-\frac{|\delta v|^2}{2}+\delta u\delta v^*\right]. 
 \label{kernelmicro2b}
\end{equation}
Indeed it is identical with the kernel~(\ref{kerneldudvbulk}) in the bulk of the spectrum since the origin is not a distinguished point for $M=1$.
For $M=2$ the Meijer G-function and hypergeometric function reduce to a  $K$-Bessel and  $I$-Bessel function respectively, which agree with the
kernel found previously in a two-matrix model describing the low-energy limit of QCD \cite{Osborn}, with large chemical potential, see e.g. \cite{APS} as well as \cite{KSi}.

Using the expression \eqref{kernelmicro2a} the microscopic level density at the origin is given by
\begin{equation}\label{densitymicro2}
\rho^{(\mu,M)}_{\rm origin}(\delta z)=K^{(\mu,M)}_{\rm origin}(\delta z,\delta z)  =\frac{1}{\pi}G^{M,\,0}_{0,\,M}\left(\mbox{}_{0,\ldots,0}^{-} \bigg| \,|\delta z|^2\right)\, _0F_{M-1}\left(\mbox{}_{1,\ldots,1}^{-} \bigg| \,|\delta z|^2\right) \ .
\end{equation}
Note that this rescaled
density has a logarithmic singularity
\be
\lim_{|\delta z|\to0}\ \rho^{(\mu,M)}_{\rm origin}(\delta z)\approx
{\rm ln}^{M-1} |\delta z|
\ee
at the origin for $M>1$. Indeed the following two limits hold:
\bea
&&\lim_{|\delta z|\to0}\, _0F_{M-1}\left(\mbox{}_{1,\ldots,1}^{-} \bigg| \,|\delta z|^2\right)\to1\ ,\nn\\
&&\lim_{|\delta z|\to0}G^{M,\,0}_{0,\,M}\left(\mbox{}_{0,\ldots,0}^{-} \bigg| \,|{\delta z}|^2\right)\approx ({\rm ln}^{M-1} |\delta z|^2)/(M-1)!\ .
\eea
For $M=1$ however this singularity is absent, as can be seen from eq.~(\ref{kernelmicro2b}).
Thus the level density takes a constant value at the origin, $\rho^{(\mu,M=1)}(\delta z)=1/\pi$, as in \cite{KS} and the Ginibre ensemble \cite{Ginibre}.

We emphasize and briefly comment on the case of more general truncations $L_j$ from appendix~\ref{sec2.3}. Due to the lack of the corresponding large $N$ expressions for the truncated series (\ref{partmacro1b}) and one-point weight (\ref{weightmacro2a}) an expansion is non-trivial. However, from universality arguments we expect that both bulk and edge fluctuations should also match with the universal Ginibre results in this case as it is the case for the fluctuations at the origin.


\subsection{Weak non-unitarity: \boldmath $N\rightarrow \infty$ and $L$ fixed}\label{sec3.1}

In this section we consider a limit called weak non-unitarity limit
introduced and studied for $M=1$ in \cite{KS},
where $N$ is taken to infinity while $L$ is kept fixed.
For fixed $L$ many things simplify since the matrices $X_j$ and their product $X^{(M)}$
are almost unitary in the limit $N\rightarrow \infty$. Indeed
we shall see that almost the whole eigenvalue spectrum of $X^{(M)}$ is concentrated in the vicinity of the unit circle in the complex plane. The macroscopic density becomes a delta function of the modulus as shown in the next subsection, whereas nontrivial universal correlations will be found when studying the inner edge of the unit circle in subsection~\ref{sec3.1.2b}.

\subsubsection{Macroscopic regime}\label{sec3.1.2}

The macroscopic level density can be readily derived via its moments
\begin{equation}\label{largeNmom1}
 \langle z^a z^{*\,b}\rangle=\frac{1}{N}\int\limits_{\mathbb{C}} z^a z^{*\,b}R_1^{(N,L,M)}(z)d^2z=\frac{\delta_{ab}}{N}\sum\limits_{j=0}^{N-1} \binom{L+j}{j}^{M} \binom{L+j+a}{j+a}^{-M},
\end{equation}
see Eqs.~\eqref{ts}, \eqref{1weight.4}, \eqref{leveldens}, and \eqref{moments}. The large $N$ asymptotics ($L$ is fixed) of these moments is given by 
\begin{equation}\label{largeNmom2}
 \langle z^a z^{*\,b}\rangle\approx \delta_{ab}\int\limits_0^1 \left[\frac{\Gamma(L+Ny+1)\Gamma(Ny+a+1)}{\Gamma(Ny+1)\Gamma(L+Ny+a+1)}\right]^M dy\approx\delta_{ab}.
\end{equation}
These moments correspond to a level density uniformly distributed on the complex unit circle,
\begin{equation}\label{densitymacLfixed}
 \rho^{(L,M)}(z)=\frac{1}{\pi}\delta(1-|z|^2).
\end{equation}
This result is indeed also true for the general case of a product of random matrices originating from different truncations, see appendix~\ref{sec2.3}.

Let us visualize what the meaning of the quite straightforward result~\eqref{densitymacLfixed} is. For $|z|$ of order unity and for large $N,L$ the density can be well approximated by the expression
\begin{equation}
 \rho^{(N,L,M)}(z)   =\frac{1}{N} K^{(N,L,M)}(z,z)
\approx
\frac{L}{\pi N M} \frac{|z|^{2/M-2}}{(1-|z|^{2/M})^2} \Theta_N \left(\left(\frac{N}{N+L}\right)^{M/2}-|z|\right),
 \label{rho_theta_N}
\end{equation}
which can be obtained from (\ref{densitymacro2}) by replacing $\alpha$ by $L/N$ and
$\mu$ by $N/(N+L)$.  Here $\Theta_N(x)$ is a sigmoidal function that changes between
zero and one in a narrow crossover region of width $\sim 1/\sqrt{N}$. In the limit $N\rightarrow \infty$ it approaches the step function $\Theta(x)$. The exact form of this function depends on $N,L,M$ but it is not essential for the argument that we are going to give below. In the first order approximation one can think of $\Theta_N(x)$  as of the step function $\Theta(x)$. For fixed $L,M$ the limit $N\rightarrow \infty$  is non-uniform and has to be taken carefully. If one takes the limit point-wise one obtains $\lim_{N\rightarrow \infty} \rho^{(N,L,M)}(z) = 0$ which is of course wrong
since the integral of the eigenvalue density has to be normalized. In fact the formula~(\ref{rho_theta_N}) gives the correct normalization
\begin{equation}
\int \rho^{(N,L,M)}(z) d^2 z  \approx \frac{L}{\pi N M} \int_0^{\left(\frac{N}{N+L}\right)^{M/2}} \frac{r^{2/M-2}}{(1-r^{2/M})^2} 2\pi r d r  =1 \ .
\label{norm1}
\end{equation}
To resolve this discrepancy it is instructive to calculate the integral of the density $\rho^{(N,L,M)}(z)$ over a disk with radius slightly smaller than one, for example
$r=(1 - c N^{-1/2})^{M/2}$ where $c$ is a positive constant. When $N$ is large enough this radius is smaller than the cut-off  $R=(N/(N+L))^{M/2}$ in the Heaviside-distribution
in eq.~(\ref{rho_theta_N}) so the disk lies entirely inside the support of the eigenvalue density. While choosing the constant $c$ one should also take into account that the function $\Theta_N$ in (\ref{rho_theta_N}) does not have a sharp threshold at $R$, but rather a sigmoidal one that extends on an interval $[R-\sigma N^{-1/2},R+\sigma N^{-1/2}]$,
which represents a smeared cut-off. One should choose $c>\sigma$ to keep the disk radius smaller than the lower value of the smeared cut-off to avoid interference with the finite $N$ boundary effects. For all points inside such a disk $\Theta_N$ in equation (\ref{rho_theta_N}) can be replaced by one and the fraction of eigenvalues inside the disk is given for large $N$ by the following integral
\begin{equation}
p \approx \frac{L}{\pi N M} \int_{|z| \le \left(1 -c N^{-1/2}\right)^{M/2}} \rho^{(N,L,M)}(z) d^2 z  =
\frac{L}{\pi N M} \int_0^{\left(1 -c N^{-1/2}\right)^{M/2}} \frac{r^{2/M-2}}{(1-r^{2/M})^2} 2\pi r d r  \ .
\end{equation}
Changing the integration variable to $x=r^{2/M}$ we find that for $N\rightarrow \infty$
\begin{equation}
p \approx \frac{L}{N} \int_0^{1 -c N^{-1/2}} \frac{dx }{(1-x)^2}
=\frac{L}{N} \left( c^{-1} N^{1/2} - 1\right) \longrightarrow 0 \ .
\label{p0}
\end{equation}
The fraction of eigenvalues inside the disk of radius $r=(1 - c N^{-1/2})^{M/2} \approx 1 - (cM/2)/N^{1/2}$ tends to zero as $1/N$ for $N\rightarrow \infty$
and the disk radius becomes one. This means that the whole interior of the disk contains almost no eigenvalues and all of them are squeezed in a narrow strip around the unit circle.
In the limit $N\rightarrow \infty$ the disk becomes an open disk. It contains no eigenvalues as shown above (\ref{p0}) while the integral over the whole disk of radius one contains all eigenvalues (\ref{norm1}). This means that all eigenvalues condense on the unit circle for $N\rightarrow \infty$ and hence
\begin{equation}
\lim_{N\to\infty}\frac1N K^{(N,L,M)}(z,z)=\rho^{(L,M)}(z) =\frac{1}{\pi}\delta(1-|z|^2),
\end{equation}
cf. the result~\eqref{densitymacLfixed}. For this reason we will only zoom into the vicinity of the unit circle in the next subsection. In the microscopic limit it can be shown that the correlations in the bulk and at the origin of the complex unit disk are highly suppressed, too.

\subsubsection{Edge fluctuations}\label{sec3.1.2b}

Consider a point on a unit circle $z=\e^{\imath \phi}$. Due to the rotational symmetry we can choose $z=1$ ($\phi=0$) without loss of generality. We are interested in eigenvalue correlations measured at two points in the vicinity of $z=1$
\begin{equation}
u = 1 - \frac{\delta u}{N} , \quad {\rm and} \quad  v = 1 - \frac{\delta v}{N},
\label{dudv}
\end{equation}
where $|\delta u|$ and $|\delta v|$ are of order unity. As usual, the first step of the calculation is to determine the large $N$ behavior of the truncated series (\ref{ts}) with argument
\begin{equation}
uv^* = 1 - \frac{1}{N} \left(\delta u + \delta v^*\right) + \frac{1}{N^2}
\delta u \delta v^* \ .
\end{equation}
We employ the representation~\eqref{partrep2}. Changing variables in eq.~\eqref{partrep2} $y_i = 1 - t_i/N$, where the $t_i$'s are of order one, leads to the following asymptotic expression:
\begin{equation}
T^{(N,L,M)}(uv^*) \approx
\frac{(-N)^{ML} N}{(L!)^M}\left. \left(\prod\limits_{j=1}^M\frac{\partial}{\partial t_j}\right)^L \left(\frac{1-\exp\left[-(t_1+\cdots+t_M)\right]}{t_1+ \cdots + t_M}\right)\right|_{t_j={(\delta u + \delta v^*)/M}}  .
\label{diffT}
\end{equation}
Hereby we used the following approximation formulae for large $N$ :
$y_1\cdots y_M \approx 1 - (t_1+\ldots +t_N)/N$, $y_i^{N+L+1} = (1 - t_i/N)^{N+L+1} \approx \e^{-t_i}$, and $\left(uv^*\right)^{1/M} \approx 1 - \left(\delta u + \delta v^*\right)/(NM)$ neglecting $O(N^{-2})$ terms. The differential operator in eq. (\ref{diffT}) acts on the function which is effectively a function of $t=t_1+\ldots+t_M$. Changing the derivatives in this operator to $\partial/\partial t_i = (\partial t/\partial t_i) \partial/\partial t = \partial/\partial t$ we obtain
\begin{equation}
T^{(N,L,M)}(uv^*) \approx
\frac{N^{ML+1}}{(L!)^M}\left. \left(-\frac{\partial}{\partial t}\right)^{ML} \left( \frac{1-\e^{-t}}{t}\right) \right|_{t={(\delta u + \delta v^*)}/M}.
\label{Tedge1}
\end{equation}
This formula cannot be easily extended to a general truncation of the matrices in the product, see appendix~\ref{sec2.3}, because of the reason discussed in the paragraph after eq.~\eqref{partrep4}. Nevertheless we expect that the simple replacement $ML\to L_1+\ldots+L_M$ and $(L!)^M\to L_1!\cdots L_M!$ should do the job.

Let us again switch to the calculation of the one-point weight.
Starting from eq.~\eqref{1weight.2} the weight reads
\begin{eqnarray}
w_M^{(L)}(u) & \approx & \frac{L^M}{\pi}\int\limits_{\mathbb{R}_+^{M}}
\delta\left(\sum_{m=1}^{M}\vartheta_m-\frac{1}{N} \delta_{R} u\right) \prod\limits_{i=1}^{M}\left(1-\e^{-\vartheta_i}\right)^{L-1}d\vartheta_i
\end{eqnarray}
 in the weak non-unitarity regime, where we introduced an abbreviation $\delta_R u = \delta u + \delta u^* = 2\Re e\delta u$.
We change the integration variables to $\theta_i$ given by $\vartheta_i= \frac{1}{N}  \delta_R u \,\theta_i$. For large $N$ the expansion\\ $\e^{-\vartheta_i} = 1 - \delta_R u \, \theta_i/N$ is sufficient up to $O(1/N^2)$ corrections. We obtain
\begin{eqnarray}
w_M^{(L)}(u) &\approx &\frac{L^M}{\pi}\Theta(\delta_R u)\left(\frac{1}{N}\delta_R u\right)^{ML-1}\int\limits_{[0,1]^{M}}\delta\left(\sum_{m=1}^{M}\theta_m-1\right) \prod\limits_{i=1}^{M}\theta_i^{L-1}d\theta_i\nonumber\\
&=&\frac{L^M N^{1-ML}}{\pi}\Theta(\delta_R u)\left(\delta_R u\right)^{ML-1}\prod\limits_{j=1}^{M-1}\left(\int\limits_0^1\theta^{L-1}(1-\theta)^{jL-1}d\theta\right),
\end{eqnarray}
and finally
\begin{equation}
\label{weightedge1}
w_M^{(L)}(u) \approx \frac{(L!)^M N^{1-ML}}{(ML-1)!\pi}\left(\delta_R u\right)^{ML-1}\Theta(\delta_R u) \ .
\end{equation}
This result can indeed be readily extended to a product of random matrices originating from different truncations $N+L_j\to N$, see appendix~\ref{sec2.3}, by replacing $ML\to L_1+\ldots+L_M$ and $(L!)^M\to L_1!\cdots L_M!$. This is the reason why we expect the same generalization of the asymptotic result~\eqref{Tedge1} for the truncated series $T^{(N,L,M)}(uv^*) $.

Collecting both asymptotic formulae (\ref{Tedge1}) and (\ref{weightedge1}) we find
\begin{equation}
\begin{split}
K^{(L,M)}_{\rm weak}(\delta u,\delta v) & \equiv \lim_{N\rightarrow \infty}
\frac{1}{N^2} K^{(N,L,M)}\left(u,v\right) \\
& = \frac{\Theta(\Re e\delta u)\Theta(\Re e\delta v)}{\pi(ML-1)!}\big(4(\Re e\delta u) (\Re e\delta v)\big)^{(ML-1)/2}
\left. \left(- \frac{\partial}{\partial t}\right)^{ML} \left( \frac{1-\e^{-t}}{t}\right) \right|_{t={(\delta u + \delta v^*)/M}} \ .
\end{split}
\label{kerneledge1}
\end{equation}
The factor $1/N^2$ results from the change of variables in the microscopic limit~(\ref{dudv}) which introduces $N^{-2}$ to the Jacobian $ d^2\delta u d^2\delta v=N^{-4} d^2ud^2v$. Note that $u$ and $v$ are complex variables.
The limiting eigenvalue density~(\ref{leveldens}) is
\begin{equation}
\rho^{(L,M)}_{\rm weak}(\delta u)=
K^{(L,M)}_{\rm weak}(\delta u,\delta u)
 = \frac{\Theta(\Re e\delta u) \left(2\Re e\delta u \right)^{ML-1}}{\pi(ML-1)!}
\left. \left(- \frac{\partial}{\partial t}\right)^{ML} \left( \frac{1-\e^{-t}}{t}\right) \right|_{t=2(\Re e\delta u)/M} \ .
\label{densityedge1}
\end{equation}
This result agrees with the one found in Ref.~\cite{KS} for $M=1$. Note that the correlation functions that follow from eq. (\ref{kerneledge1}) also agree for $M=1$ with the correlation functions found in \cite{FK} for non-Hermitian finite rank perturbations of rank $L$ of GUE matrices.

As a check one can show that the bulk fluctuations in the strong non-unitarity limit can be recovered from those of the weak non-unitarity regime by applying the following rescaling 
\be
\delta u\to Lr_0+\sqrt{L}\delta \hat{u} \ \ \mbox{and}\ \ \delta v\to Lr_0+\sqrt{L}\delta \hat{v}\ ,\ \ r_0\in\mathbb R
\ee
to eq.~(\ref{kerneledge1}) and taking the limit of large $L$. In more detail,  the exponential factor can be dropped in the $t$-derivatives of eq.~(\ref{Tedge1}) since it is highly suppressed. We obtain
\be
\left. \left(-\frac{\partial}{\partial t}\right)^{ML}\left( \frac{1-\exp\left[-t\right]}{t}\right)\right|_{t={(\delta u + \delta v^*)/M}}\approx (ML)! \left.\frac{1}{t^{ML+1}}\right|_{t={(\delta u + \delta v^*)/M}}\ .
\ee
The expansion of the $ML$-th power in the weight~(\ref{weightedge1}) is straightforward. Retaining terms up to second order in $1/\sqrt{L}$ we arrive at
\bea
\lim_{L\to\infty}K^{(L,M)}_{\rm weak}(\delta u,\delta v) &\approx& \frac{M}{4\pi Lr_0^2}
\exp\left[ -\frac{M}{8r_0^2}(|\delta\hat{u}|^2+ |\delta\hat{v}|^2- 2\delta\hat{u}\delta\hat{v}^*)\right]\nn\\
&&\times \exp\left[ -\imath\frac{M\sqrt{L}}{2r_0}\Im m(\delta\hat{u}-\delta\hat{v})
+\imath\frac{M}{8r_0^2}\Im m((\delta\hat{u})^2-(\delta\hat{v})^2)
\right].
\eea
The phase factor in the last line cancels in the correlation functions. What remains is
the Ginibre kernel with some scale factor. This factor is proportional to the macroscopic level density
\be
\rho^{(\mu,M)}(z=1-r_0)\approx \frac{M}{4 \pi r_0^2}\ ,
\ee
that is obtained from eq.~(\ref{densitymacro2}) by expanding it in $r_0$ for $\alpha=1$. The reason for $\alpha=1$ lies in the fact that the ratio $L/N$, cf. its definition~\eqref{defmualp}, does not make any sense in the scaling limit $N\gg L\gg1$. Thus one has to omit this term.

\sect{Conclusions}\label{sec4}

We have studied the complex eigenvalues of products of truncated unitary random matrices. These are obtained by removing rows and columns from larger unitary matrices of the same size or of different sizes. Our investigations generalize previous works on spectral properties of a single truncated unitary matrix \cite{KS} as well as of products of random matrices taken from the Ginibre ensemble \cite{ABu}. We mainly concentrated on the question, whether in the large matrix size limit the local spectral fluctuations also called microscopic limit lead to the same universal correlations know from non-Hermitian random matrix theory, namely the Ginibre ensemble, or if they give rise to new universality classes. This study adds an important part to the full picture of the spectral properties of product matrices consisting of a finite number of matrices recently discussed in several works \cite{ABu,AStr,Jesper,ARRS,peter2,IK,AKW,AIK,Lun,KZ}.

Our strategy was to first derive exact results for products of $M$ matrices of size $N$, resulting from truncations of unitary matrices of size(s) $N+L$ (or $N+L_j$) by $L$ (or $L_j$, with $j=1,\ldots,M$), for all parameters finite. Those results are in agreement with the very recently published works~\cite{ARRS,IK}. In the next step two large $N$ limits were identified, namely when both $N$ and $L$ become large with $L/N$ finite, named strong non-unitarity, and when $L$ remains finite at large $N$, named weak non-unitarity. We emphasize that the large $N$ limits were not done before and are the main focus of our investigations.

In the strong non-unitarity limit, the complex eigenvalues become concentrated in a supporting disk inside the unit circle (the support for unitary matrices). Three regimes were separately analyzed. At the edge and in the bulk of the support we found the same universal spectral correlations as for the Ginibre ensemble (equivalent to a single truncated unitary matrix at $M=1$) \cite{ameur,berman,TV,Mehta,FH,EK03}. The same feature was previously found for products of $M$ matrices of the Ginibre ensemble \cite{ABu}. At the origin the same microscopic correlation functions as for $M$ products of Ginibre matrices \cite{ABu} were derived. Therefore these classes labelled  by $M$ are also universal. Note that for a single Ginibre matrix with $M=1$ the origin is not special and displays the same local spectral statistics as in the bulk.

At weak non-unitarity the majority of complex eigenvalues condense on the unit circle, just as for unitary matrices. Nevertheless, microscopically they still spread out inside the complex unit disk, with a width of the order $1/\sqrt{N}$. Zooming into these edge fluctuations a new class of correlations labelled again by $M$ were found, generalizing previous results for a single truncated unitary matrix \cite{KS}.

It remains to be shown how much of this picture changes when considering different truncations $L_j$ in the large $N$ limit. Only at strong non-unitarity at the origin and in the weak non-unitarity limit we could show that the same class of universal correlations persists compared to using the same truncation $L$ for all matrices. Such investigations are currently under way. It would also be very interesting to try to use our findings in various applications mentioned in the introduction.\\

\paragraph{Acknowledgements}

We acknowledge partial support by the SFB$|$TR12 ``Symmetries and Universality
in Mesoscopic Systems'' of the German research council DFG  (G.A.), by the Alexander von Humboldt Foundation (Z.B. and M.K.),
by the Grant No. DEC-2011/02/A/ST1/00119 of National Centre of Science in Poland (Z.B.) and by the Japan Society for the Promotion of Science (KAKENHI 20540372 and 25400397) (T.N.). Z.B. especially thanks Bielefeld University for its hospitality.

\begin{appendix}

\sect{The measure for truncated unitary matrices}\label{app1}

We briefly derive the measure for the ensemble of $N\times N$ matrices obtained by truncation of a single unitary matrix $U$ distributed with respect to the Haar measure $d\chi(U)$ of $\U(N+L)$. The truncation selects the upper left block $X$ which we choose for simplicity to be quadratic\footnote{The more general result readily follows and is quoted at the end of this section.} of size $N\times N$ from a unitary matrix of size $(N+L)\times (N+L)$
\begin{equation}\label{unitary1}
U=\left(\begin{array}{cc} X & W \\ V & Y \end{array} \right) \ .
\end{equation}
The probability measure for $X$ is obtained by integrating over the remaining blocks: $V$ of size $L \times N$, $W$ of size $N\times L$ and $Y$ of size $L\times L$. It is convenient to represent the Haar measure for $U$ as a flat measure with a matrix Dirac delta function which selects unitary matrices $d \chi(U) = \delta (\eins_{N+L} - UU^\dagger ) d[U]$. We have
\begin{equation}\label{unitary}
UU^\dagger =\left(\begin{array}{cc}
XX^\dagger  + WW^\dagger &  XV^\dagger + WY^\dagger \\
YX^\dagger  + YW^\dagger &  VV^\dagger + YY^\dagger
\end{array} \right)
\end{equation}
and $d[U] = d[X] d[Y] d[W] d[V]$, to be inserted in
\begin{equation}\label{defHaar}
d \chi(X)  = \int \delta\left(\eins_{N+L} - UU^\dagger \right) d[U]\ .
\end{equation}
This result is equivalent with lemma 7 in~\cite{ARRS} proven by a QR-decomposition.

The integration over $Y$, $W$ and $V$ is performed in two steps. First we integrate out $V$ and $Y$, by introducing
an integral representation of the Dirac delta function for $(N+L)\times (N+L)$ Hermitian matrices
\begin{equation}
\delta(A) = \frac{1}{\pi^{N^2} 2^N} \int \exp\left[\imath\tr A H\right] d[H] = \frac{1}{\pi^{N^2} 2^N} \int \exp\left[\tr A(\imath H-\eins_{N+L})\right] d[H]  \ ,
\label{delta}
\end{equation}
where $H=H^\dagger$ and $A=A^\dagger$. The shift introduced in the second expression is useful for ensuring the absolute convergence over the integrals in $V$ and $Y$, see the calculation below.
After subdividing $H$ in appropriate subblocks $H_j$, $j=1,2,3$, where $H_1$, $H_3$ are Hermitian, we have
\begin{eqnarray}
d\mu(X)
 &\propto&d[X]\int \exp\left[\tr\left(UU^\dagger-\eins_{N+L}\right)
 \left(\imath\left(\begin{array}{cc} H_1 & H_2 \\ H_2^\dagger & H_3\end{array}\right)-\eins_{N+L}\right)\right]d[H_1]d[H_2]d[H_3]d[V]d[W]d[Y]\nonumber\\
 &\propto&d[X]\int {\det}^{-L-N}[\imath H_3-\eins_{L}]\exp\left[\tr(XX^\dagger+WW^\dagger-\eins_{N})(\imath H_1-\eins_{N})\right]\nonumber\\
 &&\times\exp\left[\tr H_2^\dagger(XX^\dagger+WW^\dagger)H_2(\imath H_3-\eins_{L})^{-1}-\tr(\imath H_3-\eins_L)\right]\ d[H_1]d[H_2]d[H_3]d[W]\nonumber\\
 &\propto&d[X]\int \delta(XX^\dagger+WW^\dagger-\eins_{N})\
 {\det}^{-L-N}[\imath H_3-\eins_{L}]\nonumber\\
 &&\times\exp\left[\tr H_2^\dagger H_2(\imath H_3-\eins_{L})^{-1}-\tr(\imath H_3-\eins_L)\right]\
  d[H_2]d[H_3]d[W]\nonumber\\
 &\propto&d[X]\int \delta(XX^\dagger+WW^\dagger-\eins_{N})\ d[W]\ .\label{measure2}
\end{eqnarray}
In the first step we have integrated out $V$ and $Y$ which are Gaussian, leading to determinants with powers $-N$ and $-L$, respectively. In the second step we have rewritten the integral over $H_1$ as a Dirac delta function of the invariant
$XX^\dagger+WW^\dagger=\eins_N$. The Gaussian integral over $H_2$ results in an additional power of $+L$ of the same determinant and finally the integral over $H_3$ decouples and merely contributes to the global normalization constant.
The only nontrivial contribution to the measure of $X$ comes from the remaining constraint of the upper left corner in eq.~(\ref{unitary}).

The remaining integral can be computed by writing the Dirac delta function as an integral employing eq.~(\ref{delta}) and finally integrating over $W$, i.e.
\bea
d \mu(X) &\propto& d[X] \int \exp[{\tr (XX^\dagger + WW^\dagger -\eins_N)(\imath H_1-\eins_N)}]\ d[H_1] d[W] \nonumber\\
&\propto& d[X] \int {\det}^{-L}[\imath H_1-\eins_N]\exp[{\tr (XX^\dagger -\eins_N)(\imath H_1-\eins_N)}]\  d[H_1] \nonumber\\
&=& d[X] \int {\det}^{-L}[\imath H_1-\eins_N]\exp[{\tr (X^\dagger X -\eins_N)(\imath H_1-\eins_N)}]\  d[H_1].
\label{measurefin}
\eea
The last two integrals over $H_1$ are versions of the Ingham-Siegel integral as discussed in \cite{Yan}. However we have to distinguish two cases. For the case $L\geq N$ we find
\bea
d \mu(X)
&\propto& d[X] {\det}^{L-N}[\eins_N-XX^\dagger]\Theta(\eins_N-XX^\dagger)
\label{measurefinIngham}
\eea
yielding the result given at the beginning of the paper~(\ref{measurej}). Note that one can equivalently derive the result~\eqref{measurefinIngham} from eq.~\eqref{measure2} by the following change of variables $W\to\sqrt{\eins_N-XX^\dagger}W$. Exactly in this kind of derivation the condition $L\geq N$ becomes immediate. The matrix  $WW^\dagger$ has $N-L$ zero modes in the case $N>L$ such that the remaining Dirac delta function $\delta(WW^\dagger-\eins_{N})$ can never be satisfied. Thus for the case $N>L$ we have to remain with the integral representation~\eqref{measurefin}.

For completeness we mention that repeating all the steps given above one obtains exactly the same result for a rectangular truncation, that is when the block $X$ is an $N_1 \times N_2$ rectangular matrix. Denoting the complementary dimensions of truncated blocks by $L_1$ and $L_2$ fulfilling the condition $L_1-N_2=L_2-N_1\geq0$ where $N_1+L_1=N_2+L_2$ is the dimension of the original Haar distributed unitary matrix, the result reads
\begin{equation}
\begin{split}
d \mu(X) & \propto d[X] \; {\det}^{L_1-N_2}  (\eins_{N_2} - X^\dagger X) \;
\Theta (\eins_{N_2} - X^\dagger X)
\\
 & \propto d[X] \; {\det}^{L_2-N_1} (\eins_{N_1} - XX^\dagger ) \;
\Theta (\eins_{N_1} - XX^\dagger ).
\end{split}
\end{equation}

\sect{Joint probability density of the eigenvalues}\label{app2}

In this appendix we briefly recall how to integrate out the strictly upper triangular matrix $T_i$ in the integral~(\ref{W1}). We are looking for an explicit expression only depending on the elements of the diagonal matrix $Z_i$ as given in eq.~(\ref{Ww1}). Such an expression can be derived using a Schur decomposition (we pursue this idea, too) in \cite{KS,ARRS,IK}, see also \cite{ForK} for the case $L\geq N$. Here we repeat the derivation for the sake of self-consistency and completeness.

For simplicity we will drop the index $i$ in this appendix. Let us only recall that $T$ is strictly upper triangular, that is $T_{ab}=0$ for $N\geq a\geq b\geq 1$ on the diagonal and below. Furthermore we introduce a matrix $S=Z+T$ which is (not-strictly) upper triangular and becomes quite convenient in the ensuing discussion. Our goal is to calculate the following integral
\begin{equation}
I_N = 
\int {\det}^{-L}[\imath H-\eins_N]\exp[{\tr (S^\dagger S -\eins_N)(\imath H-\eins_N)}]\  d[H]d[T]
\label{IN}
\end{equation}
which is equivalent to eq.~(\ref{W1}) up to a constant. We keep an explicit index $N$ on the left hand side since we are going to do this integral recursively in the matrix size $N$.
In the calculations below we will also use reduced matrices of size $N-1$. They will be denoted by $\eins_{N-1}$,
$H'$, $T'$, and $S'$, respectively. They are obtained from $\eins_N$,
$H$, $T$, and $S$ by removing the $N$-th column and the $N$-th row. In particular we split the matrices $S$ and $H$ in four blocks
\begin{equation}
S = \left( \begin{array}{ll} S' & \vec{v} \\ \vec{0}^T & z_N \end{array} \right)\quad{\rm and}\quad H = \left( \begin{array}{ll} H' & \vec{w} \\ \vec{w}^\dagger & h_N \end{array} \right)
\end{equation}
built of $(N-1)\times (N-1)$ blocks $S'$ and $H'$ and $1\times 1$ blocks consisting of single elements $z_N\in\mathbb{C}$ and $h_N\in\mathbb{R}$. Additionally the off-diagonal blocks correspond to vertical and horizontal vectors consisting of $N-1$ elements. The elements of the vertical vectors $\vec{v}$ and $\vec{w}$ are equal to $v_j = S_{jN}$ and $w_j=H_{jN}$, $j=1,\ldots,N-1$, respectively. All elements of the horizontal vector $\vec{0}^T$ are equal to zero. Using these conventions we can reduce the matrix dimension from $N $ to $ N-1$ in the integral (\ref{IN})
\begin{eqnarray} \label{der.3.1}
I_N&=& 
\int {\det}^{-L}\left[\imath H'-\eins_{N-1}+\frac{\vec{w}\vec{w}^\dagger}{\imath h_N-1}\right](\imath h_N-1)^{-L}\\
&&\times\exp[\tr (S^{\prime\,\dagger} S' -\eins_{N-1})(\imath H'-\eins_{N-1})+(\vec{v}^\dagger \vec{v}+|z_N|^2 -1)(\imath h_N-1)]\nonumber\\
&&\times\exp[\imath\vec{w}^\dagger S^{\prime\,\dagger}\vec{v}+\imath\vec{v}^\dagger S'\vec{w}]  d[H']d[T']d[w]d[v]d\, h_N\nonumber
\end{eqnarray}
Here we applied the following identity
\begin{equation}
\det\left[\begin{array}{cc} A & B \\ C & D \end{array}\right] =
\det\left(D\right)\ \det\left(A - BD^{-1}C\right),
\end{equation}
where in our case the first factor is the determinant of a $1\times 1$ matrix. In the first step we integrate over $\vec{v}$ such that we have
\begin{eqnarray} \label{der.3.1a}
I_N&\propto& 
\int {\det}^{-L}\left[\imath H'-\eins_{N-1}+\frac{\vec{w}\vec{w}^\dagger}{\imath h_N-1}\right](\imath h_N-1)^{-L-N+1}\exp[(|z_N|^2 -1)(\imath h_N-1)]\\
&&\times\exp\left[\tr (S^{\prime\,\dagger} S' -\eins_{N-1})(\imath H'-\eins_{N-1})-\frac{\vec{w}^\dagger S^{\prime\,\dagger}S'\vec{w}}{1-\imath h_N}\right]  d[H']d[T']d[w]d\, h_N.\nonumber
\end{eqnarray}
In the next step one can shift $H'\to H'-\imath\vec{w}\vec{w}^\dagger/(1-\imath h_N)$ which is legitimate since we do not cross any pole of the determinant. Then we end up with a Gaussian integral in $\vec{w}$ which can be done,
\begin{eqnarray} \label{der.3.1b}
I_N&\propto& 
\int {\det}^{-L}\left[\imath H'-\eins_{N-1}\right](\imath h_N-1)^{-L}\exp[(|z_N|^2 -1)(\imath h_N-1)]\\
&&\times\exp\left[\tr (S^{\prime\,\dagger} S' -\eins_{N-1})(\imath H'-\eins_{N-1})\right]  d[H']d[T']d\, h_N.\nonumber
\end{eqnarray}
The integral over $h_N$ can be performed via the residue theorem and we find the recursion
\begin{equation}\label{integralb}
I_N \propto w_1^{(L)}(z_N) I_{N-1}
\end{equation}
with
\begin{equation}
w_1^{(L)}(z_N) \propto
\left(1-|z_{N}|^2\right)^{L-1} \Theta(1-|z_N|^2) \ .
\end{equation}
Note that the function on the right hand side is independent of $N$. The only part that depends on $N$ is the coefficient which was skipped. Inserting this result into eq.~(\ref{integralb}) we eventually obtain
\begin{equation}
I_N \propto \prod_{n=1}^N w_1^{(L)}(z_n)\propto \prod_{n=1}^N  \left(1-|z_n|^2\right)^{L-1} \Theta(1-|z_n|^2)\ ,
\end{equation}
in agreement with eq.~(\ref{Ww1}) and \cite{KS,ForK,ARRS,IK}.

\sect{The model with different truncations}\label{sec2.3}

Let us consider a neat generalization where the matrices $X_j$ have still the same size $N\times N$ but they were truncated from unitary matrices of different size, $U_j\in\U(N+L_j)$, with $L_j>0$ for all $j=1,\ldots,M$. Then the measure of the product matrix $X$ reads
\begin{eqnarray}\label{matrixweight3}
d\nu(X)&=& d[X] \int {\det}^{-L_j}[\imath H_1-\eins_N]\exp[{\tr (X^\dagger X -\eins_N)(\imath H_1-\eins_N)}]\  d[H_1]\\
&=&\prod\limits_{j=1}^M{\det}^{L_j-N}(\eins_{N}-X_j^\dagger X_j)\Theta(\eins_{N}-X_j^\dagger X_j)d[X_j]\nonumber,
\end{eqnarray}
where the second line only applies in the case $L_j\geq N$.
The question is what the one-point weight function will look like. To reach this goal we go along the same lines as in the previous  subsections.

Notice that the Vandermonde determinant in the joint probability density~\eqref{jpdfev2} remains unchanged since the generalized Schur decomposition still applies, such that we have
\begin{eqnarray}\label{jpdfev3}
P^{(N,L_1,\ldots,L_M,M)}(Z)&=&\frac{1}{N!}\prod\limits_{j=0}^{M}\prod\limits_{l=0}^{N-1}
\binom{L_j+l}{l}^{-1}\prod_{a<b}^N|z_a-z_b|^2\ \prod\limits_{n=1}^Nw_M^{(L_1,\ldots, L_M)}(z_n)\ ,
\end{eqnarray}
with the normalized one point weight function
\begin{eqnarray}\label{1weight.1b}
 \hspace*{-1cm}w_M^{(L_1,\ldots, L_M)}(z_n)&=&\int\limits_{\mathbb{C}^M} \delta^{(2)}\left(z_n-\prod\limits_{j=1}^M z_{jn}\right)\prod\limits_{i=1}^M\frac{L_j}{\pi}(1-|z_{in}|^2)^{L_j-1}\Theta(1-|z_{in}|^2)d^2z_{in}\ .
\end{eqnarray}
It is obviously symmetric in the indices $L_j$.
For $L_1=\ldots=L_M=L$ we have $w_M^{(L,\ldots, L)}=w_M^{(L)}$ in comparison to the former sections.

Again the one-point weight satisfies a recurrence relation
\begin{equation}\label{recursion3}
w_{M}^{(L_1,\ldots, L_M)}(z)
=2\pi \int_0^1w_1^{(L_M)}(r') w_{M-1}^{(L_1,\ldots, L_{M-1})}\left( \frac{z}{r'} \right)\frac{dr'}{r'}
\ ,
\end{equation}
as well as with other permutations of the indices $L_j$ on the right hand side which is very similar to the one in Eq.~\eqref{recursion}. The recurrence relation in terms of the Mellin transform
\begin{equation}\label{defomega2}
M_M^{(L_1,\ldots, L_M)}(s)=\int\limits_0^1\Omega_{M}^{(L_1,\ldots, L_M)}(s)x^{s-1}dx,\ {\rm with}\ \Omega_{M}^{(L_1,\ldots, L_M)}(|z|^2) = \frac{1}{\pi}w_{M}^{(L_1,\ldots, L_M)}(z)
\end{equation}
reads
\begin{equation}\label{recursion4}
M_M^{(L_1,\ldots, L_M)}(s) = M_{M-1}^{(L_1,\ldots, L_{M-1})}(s) M_{1}^{(L_M)}(s) =\prod\limits_{m=1}^M M_{1}^{(L_m)}(s)\ .
\end{equation}
The inverse Mellin transform yields the one-point weight $w_M^{(L_1,\ldots, L_M)}$. Hereby we can benefit from the known result of the Mellin transform $M_{1}^{(L_j)}$, cf. eq.~\eqref{Mellintrafow1}. Thus the one-point weight is
\begin{eqnarray}\label{1weight.5}
 w_M^{(L_1,\ldots, L_M)}(z)&=&\frac{1}{\pi}\int_{\mathcal{C}} \prod\limits_{l=1}^M L_l!\ \frac{\Gamma(-u)}{\Gamma(L_l-u)}|z|^{2u} \frac{du}{2\pi\imath}\ \Theta(1-|z|)\\
 &=&\frac{1}{\pi}\prod_{l=1}^M L_l!\
G^{M,\,0}_{M,\,M}\left(\mbox{}_{0,\ldots,0}^{L_1,\ldots,L_M} \bigg| \,|z|^2\right)\ \Theta(1-|z|)\ .\nonumber
\end{eqnarray}
Indeed our former result~\eqref{1weight.4} can be easily retained by setting $L_1=\ldots=L_M=L$. Also the moments of this weight can be easily computed and we find
\begin{eqnarray}\label{moments2}
 \int\limits_{\mathbb{C}}w_M^{(L_1,\ldots, L_M)}(z) |z|^k d^2z&=&\prod\limits_{m=1}^M\binom{L_m+k/2}{k/2}^{-1} ,
\end{eqnarray}
which is a  generalization of Eq.~\eqref{moments}.
The corresponding kernel from which the $k$-point correlation functions are built up reads
\begin{eqnarray}\label{kernelLs}
K^{(N,L_1,\ldots,L_M,M)}(u,v)
 &=& \sqrt{w_{M}^{(L_1,\ldots, L_M)}(u)w_{M}^{(L_1,\ldots, L_M)}(v)}\ \sum\limits_{j=0}^{N-1} \prod_{m=1}^M\binom{L_m+j}{j}\ (uv^*)^{j},
 \end{eqnarray}
cf. eq.~\eqref{kcorrel}. The joint probability density, weight function and kernel agrees with the results from \cite{ARRS,IK}.

Note that most of the discussion of the asymptotics and the universality of the kernel~\eqref{kernel} carries over or can be trivially generalized to the kernel~\eqref{kernelLs}. Only the large $N$-asymptotics in the strong non-unitarity regime where one has to zoom into the local scale of the bulk and of the soft edge is quite involved due to the non-trivial saddle point equations arising from the choice $L_1\neq L_2\neq\ldots \neq L_M$.

\end{appendix}


\end{document}